\documentclass[pdflatex,sn-mathphys-num]{sn-jnl}

\usepackage{graphicx}%
\usepackage{multirow}%
\usepackage{amsmath,amssymb,amsfonts}%
\usepackage{amsthm}%
\usepackage{mathrsfs}%
\usepackage[title]{appendix}%
\usepackage{xcolor}%
\usepackage{textcomp}%
\usepackage{manyfoot}%
\usepackage{algorithm}%
\usepackage{algorithmicx}%
\usepackage{algpseudocode}%
\usepackage{listings}%

\usepackage{amsmath,amsfonts}

\usepackage{array}
\usepackage{textcomp}
\usepackage{stfloats}
\usepackage{url}
\usepackage{verbatim}

\usepackage{tabularx,booktabs,array,ragged2e,makecell}
\usepackage[table]{xcolor}
\newcolumntype{L}{>{\RaggedRight\arraybackslash}X}

\definecolor{rowgray}{gray}{0.96} 
\definecolor{headergray}{gray}{0.85} 

\usepackage{balance}

\usepackage{tabu}                      
\usepackage{booktabs}                  
\usepackage{lipsum}                    
\usepackage{mwe}                       

\usepackage{subcaption}

\renewcommand{\orcid}[1]{}          

\begin{document}

\title[Short Title]{Gesture First, LLM-Assisted Voice Complement: Exploring Multimodal Robot 'Puppeteer' Teleoperation Via Virtual Counterpart in Augmented Reality}

\author*[1]{\fnm{Yuchong} \sur{Zhang}}\email{yuchongz@kth.se}\orcid{0000-0003-1804-6296}
\equalcont{Both authors contributed equally to this research.}

\author[1]{\fnm{Bastian} \sur{Orthmann}}\email{orthmann@kth.se}\orcid{0000-0001-8542-255X}
\equalcont{Both authors contributed equally to this research.}

\author[1]{\fnm{Shichen} \sur{Ji}}\email{jishichen96@gmail.com}

\author[1]{\fnm{Michael} \sur{Welle}}\email{mwelle@kth.se}\orcid{0000-0003-3827-3824}

\author[1]{\fnm{Jonne} \sur{Van Haastregt}}\email{jonnevanhaastregt@gmail.com}

\author[1]{\fnm{Danica} \sur{Kragic}}\email{dani@kth.se}\orcid{0000-0003-2965-2953}

\affil*[1]{\orgname{KTH Royal Institute of Technology},
  \orgaddress{\city{Stockholm}, \country{Sweden}}}

\abstract{Robot teleoperation via augmented reality (AR) offers a promising path toward more intuitive human–robot interaction (HRI). We present a head-mounted AR 'puppeteer' system in which users control a physical robot by interacting with its virtual counterpart robot using large language model (LLM)-assisted voice commands and hand-gesture interaction on the Meta Quest 3. In a within-subject user study with 42 participants performing an AR-based robotic pick-and-place pattern-matching task, we empirically compare two interaction conditions: gesture-only (GO) and combined voice+gesture (VG) on performance and user experience (UX). In VG, voice and gesture operate in a sequential role-allocated manner, with voice handling high-level navigation and gesture handling fine manipulation. Our results show that GO currently provides more reliable and efficient control for this time-critical task, while VG introduces additional flexibility but also latency and recognition issues that can increase workload. We additionally analyze how prior robotics expertise differentiates performance and UX across conditions. Based on these findings, we distill a set of design guidelines for AR 'puppeteer' metaphoric robot teleoperation, framing multimodality as an adaptive strategy that must balance efficiency, robustness, and user expertise rather than assuming that additional modalities are universally beneficial.}


\maketitle

\section{Introduction}

\begin{figure}
    \centering
    \includegraphics[width=\linewidth]{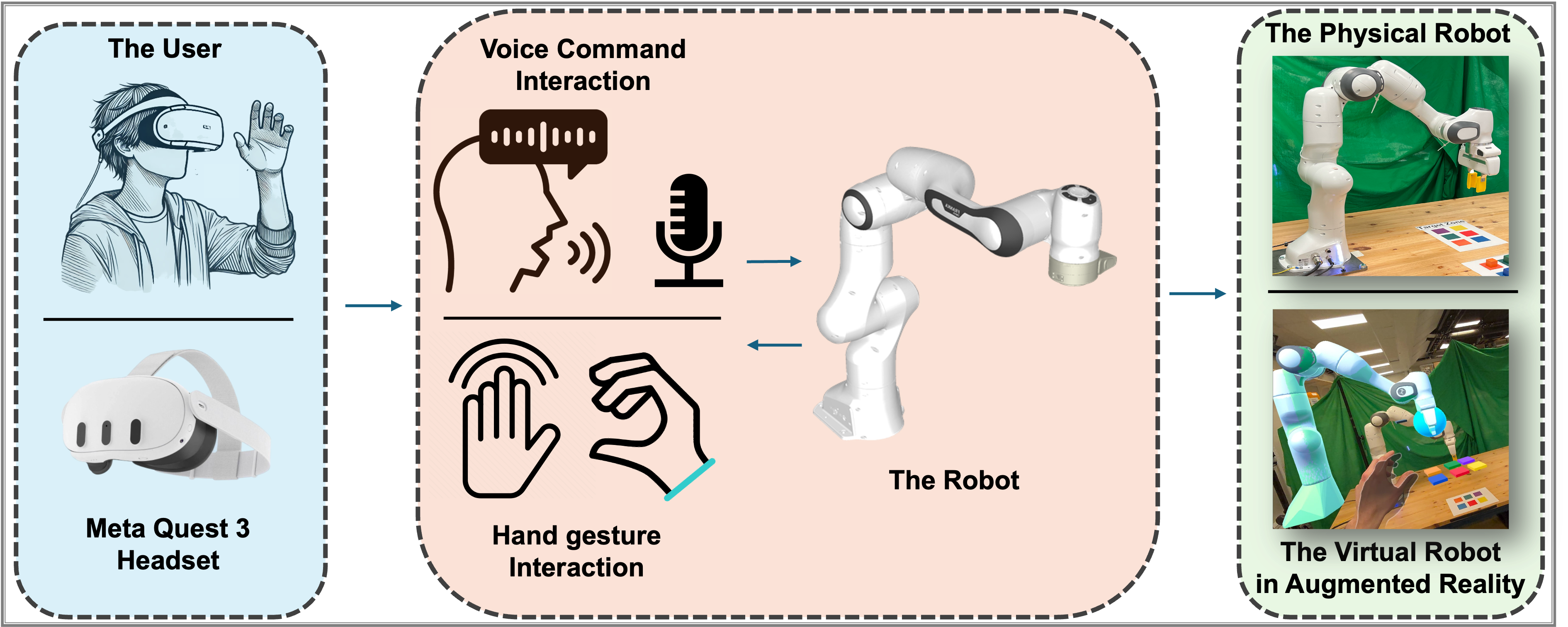}
    \caption{The AR-based multimodal robot 'puppeteer' system used in this study, enabling voice and gesture interaction for teleoperation.}
    \label{fig:teaser}
\end{figure}

Robotics is rapidly reshaping numerous industries by facilitating automation, manufacturing, assistance, and collaboration across various domains \cite{ribeiro2021robotic,hentout2019human,xia2024shaping}. As intelligent robots become increasingly integrated into daily human-centered environments such as healthcare \cite{olaronke2017state,kyrarini2021survey} and social interactions \cite{richert2016socializing,zhang2024human}, it becomes essential to develop intuitive and natural interaction mechanisms. Traditionally, human-robot interaction (HRI) seeks to establish effective mutual communication between humans and robots to enhance task performance through collaboration or to improve user experience (UX) from a human-centered perspective \cite{hopko2021effect,apraiz2023evaluation,zhang2025mind}. Augmented reality (AR)—a technology that overlays virtual elements onto the physical environment using specialized devices \cite{zhou2008trends,zhang2021novel}—has emerged as a powerful and versatile tool to enhance HRI by providing real-time feedback \cite{saeedi2018navigating,suzuki2022augmented}. Historically, AR has been extensively employed for visualization due to its capability to directly present additional information in front of users \cite{zhang2022initial,nowak2021augmented,zhang2021supporting}. By bridging the physical and digital worlds, AR significantly improves usability, intuitiveness, situational awareness, and accessibility, making it increasingly valuable in the development of effective HRI systems today \cite{yoo2025study,bambusek2019combining,gaschler2014intuitive}.

Within HRI systems, leveraging AR interfaces allows users to interact with robots in a more intuitive \cite{chong2009robot,zhang2023playing} and spatially aware manner \cite{lunding2023ar}. This approach reduces cognitive load and operational difficulties \cite{walker2018communicating,moya2023augmented} associated with traditional interaction methods. Over recent years, numerous studies have explored the use of AR in robotic teleoperation, recognizing that additional virtual information can provide essential insights leading to more informative interactions and more robust robot control \cite{arevalo2021assisting,makhataeva2020augmented}. For instance, AR has been successfully employed to improve robotic navigation efficiency \cite{mohareri2011autonomous} and enhance social interactions with dialogue-based robots \cite{barakonyi2004agents}, demonstrating the advantages offered by AR integration. Notably, most previous research on AR-assisted HRI or human-robot collaboration has predominantly focused on direct manipulation and control of physical robots. Among these contributions, the work by Haastregt et al. \cite{van2024puppeteer} stands out by introducing the first controller-based AR system for virtual robot teleoperation using a head-mounted display (HMD), enabling users to manipulate virtual robots via physical controllers.

An advancement in HRI has been the integration of multimodal interaction -- combining various channels especially gestures, voice commands, and haptic feedback -- to facilitate enhanced communication between humans and robots \cite{salem2011friendly,torta2015evaluation,okita2011multimodal}. However, the interdisciplinary field of multimodal interaction combined with AR-based virtual robotic teleoperation remains relatively unexplored. This approach is especially promising in extended reality (XR) environments, where users benefit from immersive virtual projections, improved spatial awareness, real-time feedback, and intuitive control methods when interacting with virtual robots, thus indirectly influencing physical robot systems. Compared to single-input interaction systems, multimodal interaction has the potential to narrow the gap between human intentions and robotic execution, leading to more seamless, efficient, and user-friendly interactions \cite{ali2019design,lee2013usability,zhang2023see}. The sequential role-allocated multimodal interaction, in which different modalities are assigned to distinct task stages or action types \cite{turk2014multimodal}, is a common and empirically grounded integration approach which does not require simultaneous signals \cite{oviatt1999ten}. Previous research has demonstrated that integrating multiple modalities notably enhances overall UX and usability within HRI contexts \cite{lunghi2019multimodal,lazzeri2014development,cao2023investigating}. Nonetheless, applying these multimodal approaches specifically to AR-based virtual robot control presents novel challenges, particularly concerning accurate voice command recognition, gesture detection, and synchronized robotic execution \cite{zhang2024vision}.

In this research, we investigate how multimodal interaction -- specifically voice and hand gesture -- shapes teleoperation of a physical robot via its virtual counterpart in head-mounted AR, with a focus on expertise-aware target groups. We present an AR 'puppeteer' prototype that synchronizes a physical robot and its virtual counterpart for teleoperation, integrating large language model (LLM)-assisted voice commands with hand-gesture interaction sequentially. Rather than treating virtual simulation and real-world robot control as separate workflows, our system lets users directly manipulate the virtual counterpart as a graphical control agent, translating these virtual manipulations into physical robot motion. By involving an empirical study, we aim to examine how different modalities affect task performance, usability, and user experience, and how these effects vary with users’ prior expertise. Based on observed findings, we derive actionable, expertise-sensitive design implications for multimodal AR teleoperation, and discuss the implications for the HCI/HRI community. The key contributions include:

\begin{itemize}
  \item Present an HMD AR 'puppeteer' teleoperation prototype that instantiates gesture-only (GO) and voice+gesture (VG) conditions for virtual-to-physical robot interaction.
  \item Report a within-subject user study (N=42) comparing GO and VG for a structured pick-and-place teleoperation task, analyzing task performance, usability, and UX.
  \item Derive empirical, expertise-aware design implications for sequential voice+gesture interaction for AR 'puppeteer' robotic teleoperation, contributing actionable recommendations for multimodal HRI system design.
\end{itemize}

\section{Related work}
\subsection{XR for HRI}
Over the past decades, HRI has evolved significantly through the integration of XR—the umbrella term of virtual reality (VR), AR, and mixed reality (MR) \cite{williams2018virtual,wang2024towards}. These immersive technologies provide new paradigms for interaction beyond traditional screens and controllers. As early as 1999, Freund et al. \cite{freund1999projective} demonstrated that projective VR could support fundamental robotic functions such as task planning. In subsequent years, VR has been widely applied, for example, Matsas et al. \cite{matsas2017design} proposed a VR-based training system for human-robot collaboration in industrial settings. Similarly, Murnane et al. \cite{murnane2021simulator} utilized VR environments to collect interaction data for training verbal HRI models, while Villani et al. \cite{villani2018use} employed VR as an evaluation tool to simulate complex HRI scenarios that would be difficult to reproduce in real-world environments. AR, which also encompasses MR, has gained broader attention in HRI due to its ability to overlay digital content onto the physical world, enabling real-time visualization, intuitive control, and spatial awareness. In collaborative HRI, Chacko et al. \cite{chacko2019augmented} developed a smartphone-based AR interface for pick-and-place tasks, allowing users to visually select target objects within a shared workspace. Michalos et al. \cite{michalos2016augmented} introduced an AR system for industrial co-working environments, providing assembly guidance, safety cues, and live production feedback, validated in an automotive use case. For teleoperation, Hedayati et al. \cite{hedayati2018improving} presented AR interfaces that improved visual feedback and user performance during aerial robot inspections. Walker et al. \cite{walker2019robot} further extended this by introducing a virtual surrogate drone in AR to enhance control efficiency, reduce cognitive load, and support multitasking—showing measurable improvements across both novice and expert user groups.

\subsection{Multimodal Interaction in XR for HRI}
Multimodal interaction -- leveraging a combination of input modalities such as gesture, voice, gaze, and haptics -- has become a promising area in XR-enabled HRI for enhancing control precision, accessibility, and user engagement. In early VR-based research, Savage et al. \cite{savage1998virbot} developed a VR system that enables users to simulate controlling a mobile robot in virtual environments. The robot responds to voice and simple gesture commands, with contextual reasoning enhancing command recognition. Notably, the virtual robot already adopted a sequential role-allocation between voice for navigation and contextual gesture for manipulation, establishing an early precedent. More recently, the focus has shifted to AR-based multimodal systems. Krupke et al. \cite{krupke2018comparison} developed an MR system that lets users control co-located virtual industrial robot arms for pick-and-place with comparing heading+speech versus pointing+speech for pick selection. Park et al. \cite{park2021hands} proposed a hands-free HRI method using head gestures, eye gaze, and deep learning to enable efficient robot control via indirect object manipulation and a digital twin system. Chan et al. \cite{chan2022multimodal} developed an AR platform integrating speech, gestures, gaze, electromyography, and tactile feedback for robot trajectory programming, emphasizing low-attention, force-aware interaction. Szczurek et al. \cite{szczurek2023multimodal} designed a multimodal AR interface for collaborative intervention in hazardous environments, using head-mounted displays to support shared workspaces and reliable control. More recently, Tao et al. \cite{tao2025lams} proposedan LLM-driven approach that automatically switches teleoperation control modes based on task context, demonstrating that LLMs can reduce the cognitive burden of manual mode transitions in manipulation tasks. Hu et al. \cite{hu2025gesprompt} demonstrates a related approach in VR, where co-speech gesture input is combined with LLM-based speech processing to resolve spatial ambiguities in 3D environment commands, highlighting the complementary roles of voice and gesture in XR contexts."

\subsection{Robot Puppeteer and Research Gap}
Regarding controlling robots using AR featured virtual robots, Duan et al. \cite{duan2023ar2} introduced an iOS pipeline: AR2-D2 to turn phone videos of a person manipulating objects into robot demonstrations via an AR Franka overlay and motion planning, but removing the need for a real robot during data collection and targeting behavior-cloning pipelines. Similarly, Wang et al. \cite{wang2024eve} extended the work by proposing an iOD AR app -- EVE -- to let users set and edit AR waypoints on iOS to gather training trajectories and verify them, emphasizing data quality and downstream policy results, not live control of any physical robots. Despite these advancements, research on virtual-to-physical robot control using a puppeteering metaphor in AR remains limited. Barakonyi et al. \cite{barakonyi2004agents} explored animated agents in AR to facilitate seamless control across physical and virtual domains, while Walker et al. \cite{walker2019robot} demonstrated the use of virtual drone surrogates to improve teleoperation and reduce user stress. Krupke et al. \cite{krupke2018comparison} proposed an HMD-based AR concept in which a visualized robotic arm is co-located with the real robot, allowing users to control the virtual arm to rehearse tasks before executing them on the physical robot. In terms of virtual-to-physical synchronization, van Haastregt et al. \cite{van2024puppeteer} introduced the first AR-based 'puppeteer' system that allows users to control a physical robot for simple movements via manipulation of a virtual counterpart using a physical HMD controllers in hands, while Zhang et al. \cite{zhang2025llm} upgraded similar system by introducing the controller-free interaction with simple LLM integration for voice command.

However, several gaps remain. First, the use of multimodal interaction, especially comparing voice and gesture interaction, in AR-based virtual-to-physical robot puppeteering has received limited attention. Second, there is a lack of empirical evaluations that jointly examine task performance, contextual usability, and UX in such systems. Third, prior work offers little design-oriented guidance that accounts for differences between experts/non experts. Fourth, prior work has not systematically examined sequential role-allocated voice+gesture architectures, in which modalities serve distinct task stages rather than operating
simultaneously, in robotic teleoperation with virtual counterpart with AR, despite this being a common deployment pattern in real teleoperation systems.

Our work addresses these gaps by investigating how multimodality (voice and gesture) shapes AR-based robot teleoperation in a puppeteer-like metaphor. Unlike a full digital twin, our virtual robot in AR functions as a standalone graphical proxy -- without live bidirectional data flow with the physical robot -- similar to the setups described in \cite{van2024puppeteer,zhang2025llm}, but extended with an expertise-aware, multimodal evaluation.

\section{Research design}
\subsection{The Robot 'Puppeteer' System}
\begin{figure*}[!t]
    \centering
    \includegraphics[width=\linewidth]{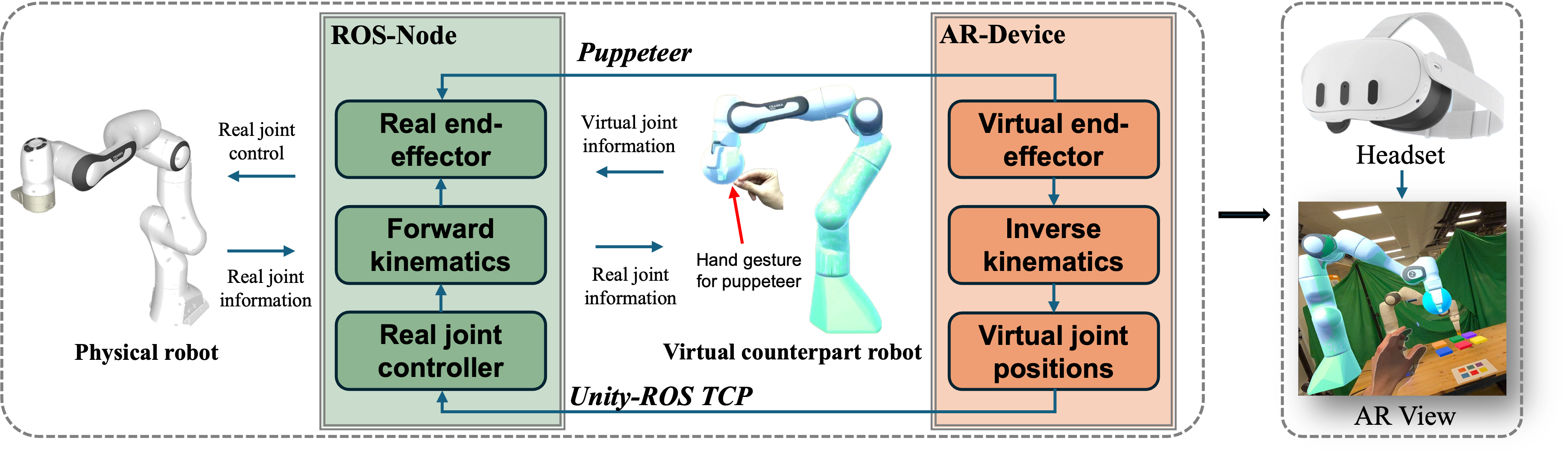}
    \caption{System realization of the AR 'puppeteer' framework. \textbf{Left: the technical implementation; right: the headset and the AR view.} The user wears an AR headset that overlays a virtual robot onto the physical workspace. By interacting with the virtual robot -- specifically manipulating its end-effector via hand gestures. The AR device computes the corresponding virtual joint positions through inverse kinematics and sends these to a ROS node. The ROS node translates these into control commands for the physical robot using a joint-level PD controller. Simultaneously, the current joint state of the physical robot is sent back to the AR device, where forward kinematics reconstructs the real end-effector pose. This allows visualization of both the desired (virtual) and actual (physical) robot states within the AR view.}
    \label{fig:imp}
\end{figure*}

The developed AR-based robotic 'puppeteer' system centers a virtual counterpart robot, which mirrors a physical robot in real-time with Meta Quest 3. Through continuous synchronization, the physical robot replicates the movements of its virtual counterpart. As illustrated in Fig.~\ref{fig:teaser}, the system is built upon the Franka robotic arm -- a widely used 7-degree-of-freedom collaborative robot designed for manipulation tasks, with a 3D-printed yellow gripper mounted at the end-effector \footnote{\href{https://franka.de/}{https://franka.de/}}. The virtual counterpart robot is rendered in light green within the AR environment and projected directly in the user's field of view to facilitate intuitive interaction. The experiment subjects used in this study include a few pre-defined zones within the joint range of the physical robot based on six colored cubes (see Section ~\ref{procedure} for details). The connection between the physical robot and the virtual robot was achieved via Unity-Robot Operating System (ROS) TCP \footnote{\href{https://github.com/Unity-Technologies/ROS-TCP-Endpoint}{https://github.com/Unity-Technologies/ROS-TCP-Endpoint}}. The technical realization of the 'puppeteer' system is displayed in Figure ~\ref{fig:imp}. The position and orientation of the virtual robot are defined by user configuration in the real study, as described in the following sections.

\subsection{The Voice and Gesture Interaction} 
The system implements sequential role-allocated multimodal interaction: voice and gesture interactions are not fused together, but are assigned to distinct task stages and action types.
Voice commands are designed for discrete, high-level navigation of the virtual robot to predefined zones, while hand gestures handle continuous, fine-grained spatial manipulation on the virtual
end-effector. This architecture reflects established role-allocated sequential principles in \textit{"exclusive"} multimodal system design \cite{turk2014multimodal}.

Voice commands are enabled through integration with an LLM. Audio input is captured via a microphone and processed in Python using RealtimeSTT, an open-source, real-time speech-to-text library developed by Kolja Beigel\footnote{\href{https://github.com/KoljaB/RealtimeSTT}{https://github.com/KoljaB/RealtimeSTT}}. Voice recognition is activated by a pre-defined wake word, “Blueberry,” detected using GPU-accelerated transcription powered by Faster Whisper, which a built-in of RealtimeSTT\footnote{\href{https://github.com/SYSTRAN/faster-whisper}{https://github.com/SYSTRAN/faster-whisper}}. The wake word was selected by the authors from a shortlist offered by Faster Whisper, based on factors such as pronunciation clarity and tonal distinctiveness. Upon successful detection of the wake word, the message “Blueberry is listening” is immediately presented as textual feedback on the upper section of the user’s HMD display, indicating that the system is actively listening. No feedback is provided if voice activation fails to detect a valid input.

Once transcription begins, users issue commands instructing the virtual counterpart robot (instantly followed by the physical robot) to navigate to a designated area, which may be associated with a specific color. The system first attempts \textit{quick detection} by matching the transcribed text against hardcoded regular expressions corresponding to pre-defined command keywords (e.g., cube color names or directional phrases such as “move towards black” or “move to black”). If no match is found, the transcribed input is passed directly to a round-robin pool of lightweight, locally hosted Llama 3.2 1B-Instruct Q6 LLMs\footnote{\href{https://huggingface.co/wirthual/Meta-Llama-3.2-1B-Instruct-llamafile}{https://huggingface.co/wirthual/Meta-Llama-3.2-1B-Instruct-llamafile}}. Upon successful interpretation—via either method—the system initiates automated movement of the end-effector to a pre-defined zone corresponding to the identified color.

\begin{figure}[t]
\centering
  \begin{minipage}[t]{0.28\linewidth}
    \centering
    \includegraphics[width=.7\linewidth]{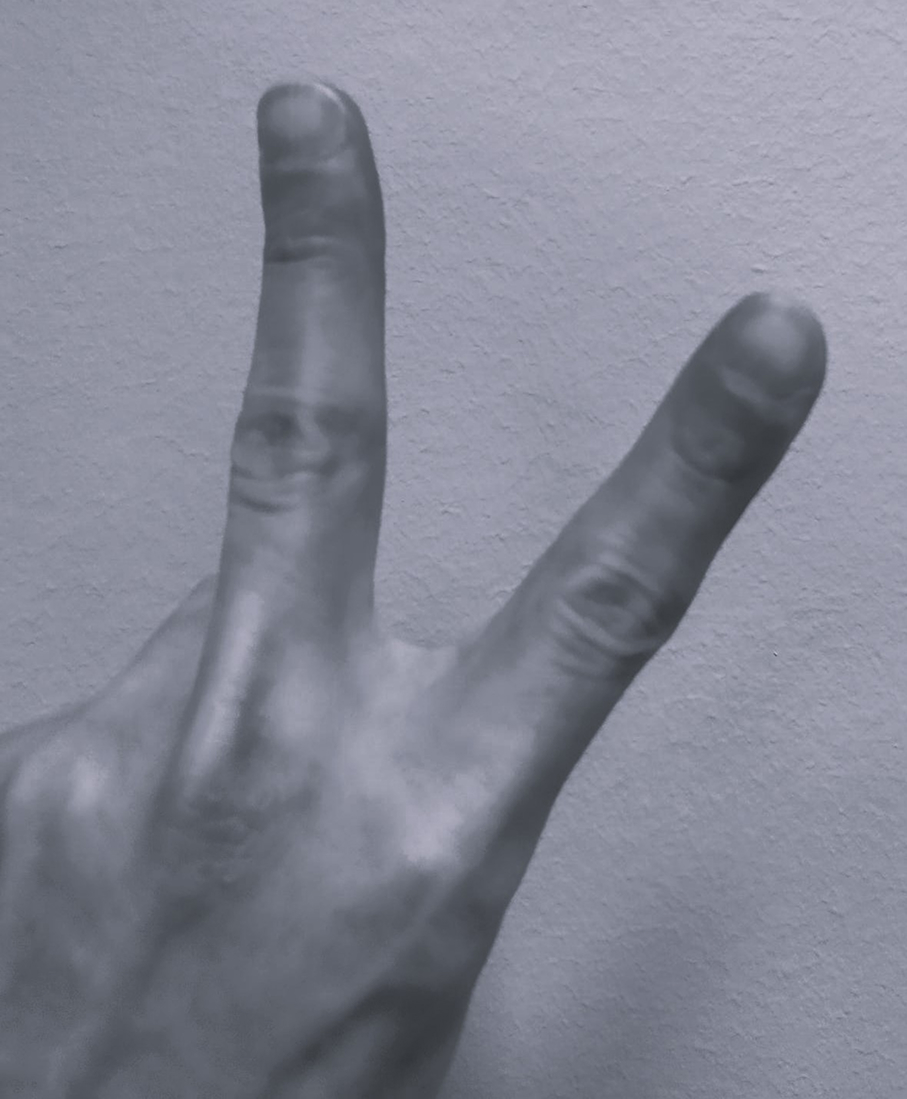}
    \subcaption{Gesture for spawning the virtual robot with 'victory' sign.}
    \label{fig:spawn action}
  \end{minipage}\hspace{0.01\linewidth}   
  \begin{minipage}[t]{0.28\linewidth}
    \centering
    \includegraphics[width=.7\linewidth]{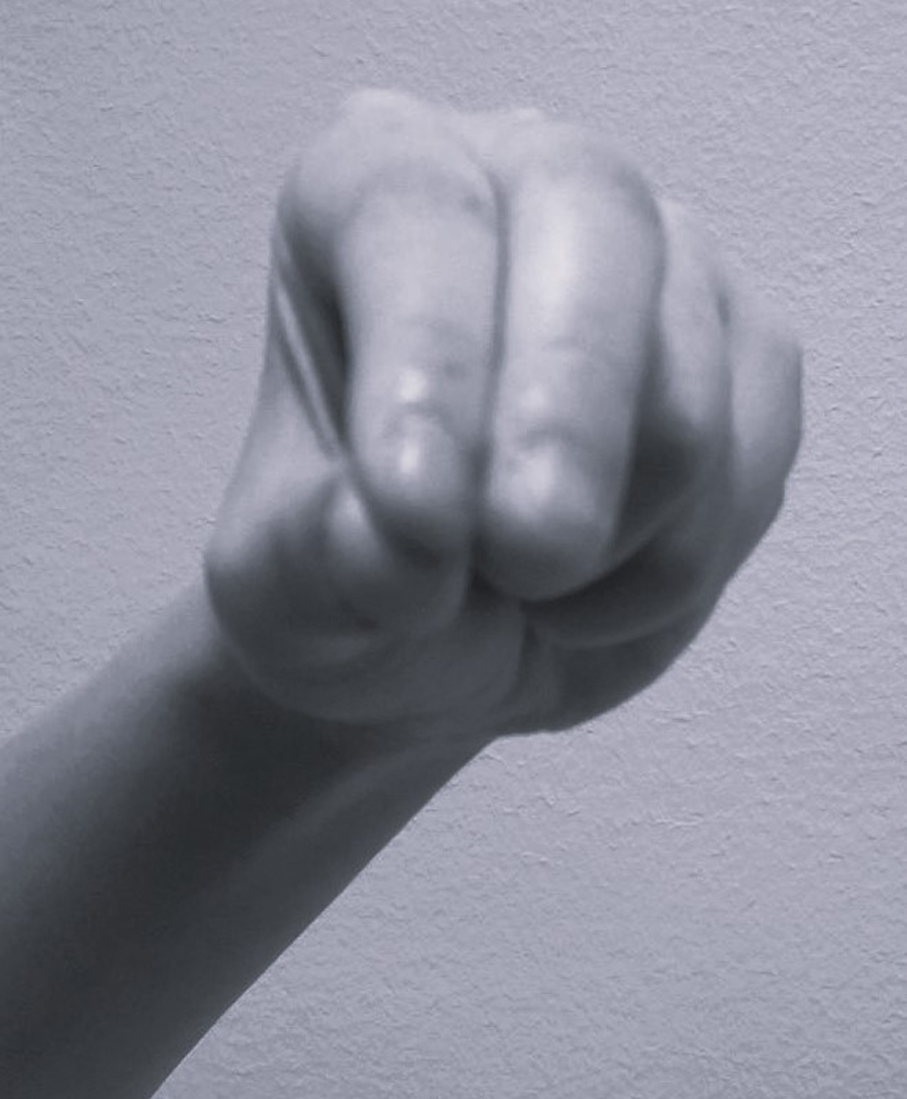}
    \subcaption{Gesture for puppeteering the virtual robot with three fingers' pinch.}
    \label{fig:puppeteer action}
  \end{minipage}\hspace{0.01\linewidth}   
  \begin{minipage}[t]{0.28\linewidth}
    \centering
    \includegraphics[width=.7\linewidth]{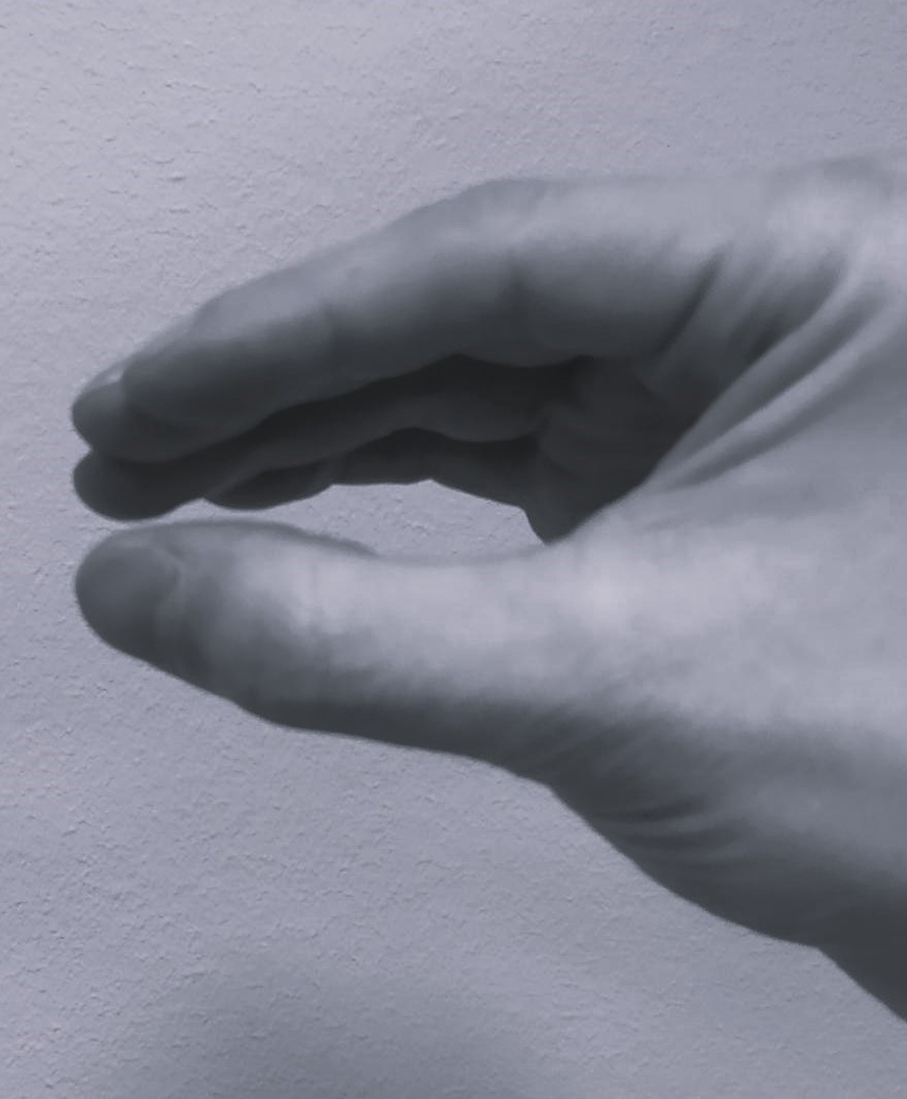}
    \subcaption{Gesture for controlling the gripper with all fingers stretched out.}
    \label{fig:gripper action}
  \end{minipage}

  \caption{Hand gestures designed for interacting with the virtual robot.}
  \label{fig:gestures}
\end{figure}

For gesture interaction, the built-in hand tracking system of the Meta Quest 3 was utilized in conjunction with the OpenXR framework in Unity. On top of this, a custom gesture interpretation system was developed to map tracking data to a set of pre-defined gestures assigned to the user’s primary and secondary hands. For right-handed users, the right hand is designated as the primary hand and the left as the secondary; this is reversed for left-handed users. Gestures for manipulating the movement of the virtual robot and activating puppeteering are linked to the primary hand, while the grip gesture is assigned to the secondary hand. Before using the system, users select their dominant (primary) hand, which automatically defines the secondary hand. The system recognizes three main gestures:

\begin{enumerate}
\item A “victory” sign, used to trigger the spawn action and instantiate the virtual robot at the gesture’s location using either hand (Fig.\ref{fig:spawn action}).
\item A three-finger pinch (thumb, index, and middle fingers touching) performed with the \textbf{primary hand} to activate the puppeteering action (Fig.\ref{fig:puppeteer action}).
\item A grasp gesture using the \textbf{secondary hand} (index, middle, ring, and pinky fingers aligned, with the thumb either touching or extended), which controls the robot’s gripper (Fig.~\ref{fig:gripper action}).
\end{enumerate}

The puppeteering gesture allows the user to interact with the virtual robot's end effector,  which was designed as a virtual green sphere surrounding it (Fig.~\ref{fig:imp}), enabling real-time movement control. To initiate this interaction, the user must maintain the pinch gesture within the bounds of the green sphere; movement is triggered only while the gesture remains active and within this region. This interaction will drive the inverse-kinematics controller which sets the joint position controller of the physical robot. The grasp gesture independently controls the opening and closing of the physical robot’s gripper and can be executed concurrently with the puppeteering gesture.

        

\subsection{Apparatus}
The experimental setup included a Meta Quest 3 XR headset, a Zoom H2N microphone mounted on a fixed stand, a dedicated PC for voice command processing, and a separate PC running ROS to control the Franka robotic arm. The voice processing PC was equipped with an NVIDIA RTX 2070 GPU, Intel i9-10900K processor, and 125 GB RAM, and hosted lightweight local LLMs for parsing speech commands. The AR application was developed using Unity 2022.3.20f1 and deployed on Android. 

\section{User study}
\subsection{Overview of the Study}
\begin{figure*}[!t]
\centering
\captionsetup[sub]{justification=centering}

\begin{subfigure}[t]{0.48\linewidth}
  \centering
  \includegraphics[width=.7\linewidth]{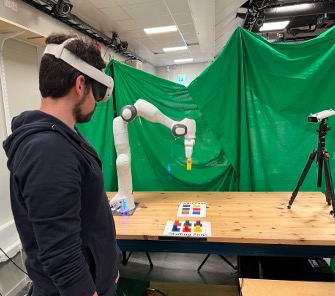}
  \caption{Example scene of the user study.}
  \label{fig:us}
\end{subfigure}\hfill
\begin{subfigure}[t]{0.48\linewidth}
  \centering
  \includegraphics[width=.7\linewidth]{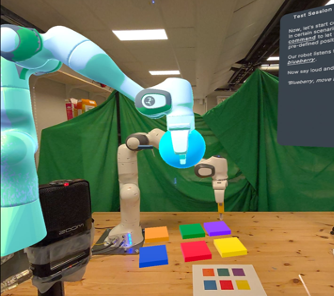}
  \caption{The user view of the system.}
  \label{fig:view}
\end{subfigure}

\medskip

\begin{subfigure}[t]{0.48\linewidth}
  \centering
  \includegraphics[width=.7\linewidth]{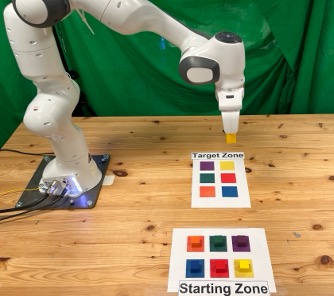}
  \caption{The setup with pre-defined zones and colored cubes.}
  \label{fig:setup}
\end{subfigure}\hfill
\begin{subfigure}[t]{0.48\linewidth}
  \centering
  \includegraphics[width=.7\linewidth]{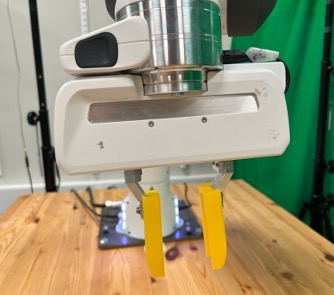}
  \caption{The gripper of the robot at the end-effector.}
  \label{fig:gripper}
\end{subfigure}

\caption{Example scenes of the setup and user study.}
\label{fig.overall_setup}
\end{figure*}

We conducted a within-subject comparative user study to evaluate the effectiveness of the multimodality in our robot puppeteering system. Participants were asked to complete two identical cube pattern-matching tasks where they were instructed to pick up six colored cubes (red, yellow, blue, orange, purple, and green) from a designated starting zone, grasp them, and place them in the target zone to match a predefined color cell (Fig.~\ref{fig:setup}). The cube pick-and-place task is a well-established benchmark in robotic teleoperation research, widely recognized for its effectiveness in assessing interaction methods \cite{fang2014novel,ostanin2020human,pandey2014towards}. Additionally, pattern-matching tasks have been commonly used in HCI and psychology studies to measure task accuracy and cognitive load \cite{seaborn2010exploring,swan2015matching,nickerson1973visual}. To better match the 'puppeteer' nature of our system and the hands-on demands of pick-and-place manipulation, we implemented two interaction conditions: gesture-only (GO) and voice + gesture (VG) (Fig.\ref{fig:us} and Fig.\ref{fig:view}). A voice-only condition was not included as it offers limited spatial precision and was deemed impractical for this task type. In the GO condition, users relied exclusively on hand gestures for task execution, while in the VG condition, they could use a combination of voice commands and gestures. The independent variable was the interaction modality (GO vs. VG), where VG was operationalized as a sequential role-allocated manner. Within this context, we initially hypothesized:

\begin{itemize}
    \item \textbf{H1:} The VG condition will lead to better task performance compared to the GO condition.
    \item \textbf{H2:} The VG condition will enhance UX and usability compared to the GO condition.
\end{itemize}

\subsection{Participants}
We recruited 42 participants, primarily from a local university, with an age range of 22 to 33 ($M=26.35, SD=3.7$). Among them, 57\% self-identified as male ($n=24, M=27.22, SD=3.58$) and 43\% as female ($n=18, M=25.18, SD=3.64$). No participants withdrew voluntarily, reported color blindness, or expressed discomfort during the study. Due to unresolvable technical issues, two participants were unable to complete the study, resulting in a final dataset of 40 valid samples. Each participant completed the experiment independently and successfully followed the study procedures. Of the 40 participants who completed the study, 24 were classified as roboticists (holding intermediate or advanced robotics knowledge with hands-on experience) and 16 as non-roboticists. 34 participants reported education level as graduate degree and doctorate degree. Regarding prior XR experience, 25 participants had previously used AR, while 29 had experience with VR. In terms of robotics expertise, 24 participants were classified as roboticists, possessing intermediate and advanced knowledge and hands-on experience in robotics. Only two participants reported as left-handed. The study was conducted in a controlled robotic laboratory, free from external distractions or background noise, ensuring a consistent and immersive study environment. Each participant received a \$10 gift card as a token of appreciation. The study duration ranged from 30 to 45 minutes per participant, including the pre-briefing, task execution, and the questionnaire filling. All data collection procedures adhered to ethical guidelines approved by the authors’ home institutions.

\begin{figure*}[!t]
    \centering
    \includegraphics[width=\linewidth]{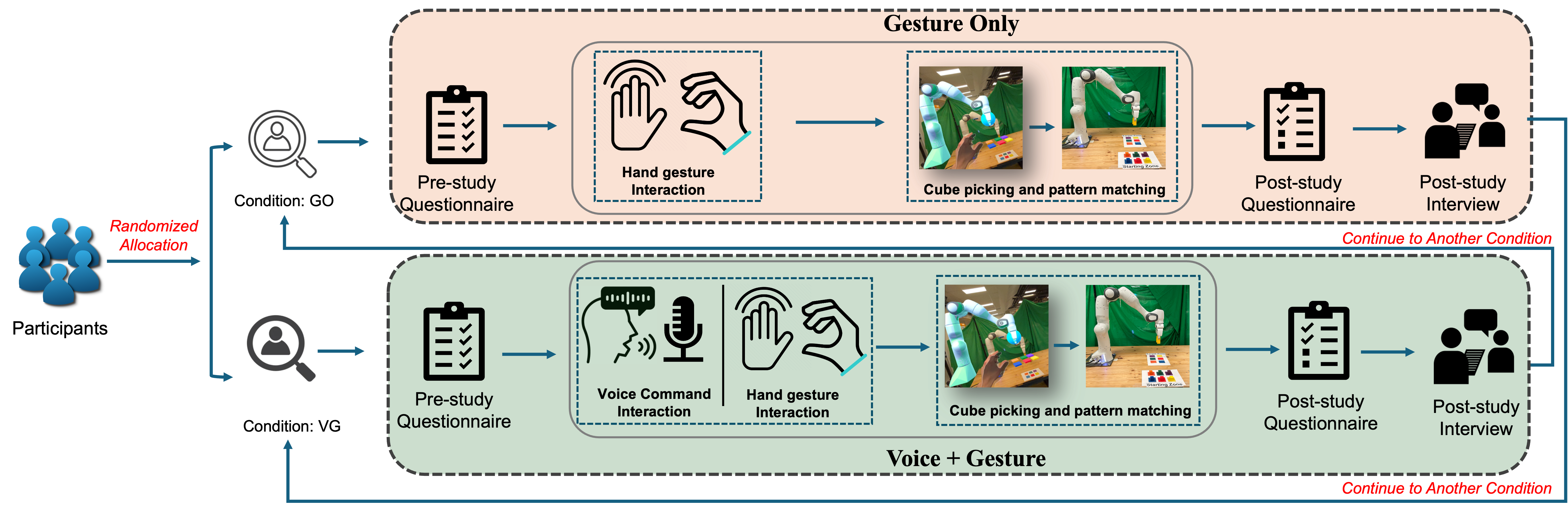}
    \caption{An overview of the user study workflow.}
    \label{fig:us_process}
\end{figure*}



\subsection{Procedure}
\label{procedure}
Before the formal study began, each participant completed a pre-testing session to validate the functionality of the system. After wearing the HMD and launching the app, participants were first asked to indicate whether they were right-handed or left-handed, and this information was recorded in the app to enable correct gestures during the interaction. Following this, they were instructed to test the voice command by speaking into the microphone using the test query (\textit{“Blueberry, move to black”} which directed the robot to a pre-defined location separate from any of the cube positions.) and hand gesture (all three gestures: spawning, moving the virtual robot, and implementing the grip action) interaction using sample queries, which closely resembled those used in the formal study. The scripted voice command vocabulary was pre-formulated to control the variability of spontaneous speech and to ensure consistent task conditions across participants in the study. This session ensured participants were comfortable with the AR interface and interaction modalities before proceeding to the main experiment. Only after confirming their readiness and demonstrating sufficient familiarity with the system did participants sign the consent form and proceed to the formal study.

Participants began by completing a pre-study questionnaire designed to collect background information, including prior experience with VR/AR technologies and robotics. They then proceeded to the task execution phase, which involved a cube pick-and-place and pattern-matching task to be completed within a 5min (300s) time limit. This time limit was set to keep total session length manageable (30–45 minutes including pre-briefing, two experimental conditions, post-study questionnaires, and the post-interview) and to reduce participant fatigue, which prior studies have shown that longer trials in tepeoperation studies can casue performance and workload measurements degradation \cite{hopko2021effect,biener2022quantifying}. This 300s setting was applied identically to both GO and VG conditions to preserve within-subject comparability, and it was communicated to participants before each trial so that the time limit formed part of the shared task framing. An example of the user’s AR view is shown in Fig.~\ref{fig:view}, where the instruction interface appears floating on the right side, the microphone on the left, and the virtual robot is positioned directly in front of the user after spawning, which remained still throughout the whole study. Participants were instructed to maintain simultaneous visual awareness of both the virtual and physical robots throughout the task. Although virtual reference cubes were displayed in the AR environment, participants were informed that these elements were intended solely for contextual reference and not for interaction. During the task, participants interacted with the virtual counterpart robot for control and manipulation, while visually focusing on the physical robot to perform the pick-and-place actions in the 'puppeteer' metaphor. The task began by moving the robot from its initial location to the starting zone, where colored cubes were placed in correspondingly colored cells (Fig.\ref{fig:setup}). Using the yellow gripper mounted on the robot’s end-effector (Fig.\ref{fig:gripper}), participants grasped each cube and placed it into the matching cell in the target zone to complete a color pattern.

Upon completing the task, participants filled out a post-study questionnaire to assess subjective impressions and usability perceptions. This was followed by a post-study interview, during which they provided qualitative feedback on their experiences with the two interaction modalities. Due to inherent system fragility, the physical robot occasionally stopped functioning when joint limits were exceeded or when movement was too rapid due to misoperations. These interruptions did not affect task continuation. A built-in timer in the app automatically paused during such events and resumed once the task was restarted by the experimenters. To ensure accurate and consistent time tracking, a secondary manual timer was also maintained by the researchers. During the study, one author was responsible for recording performance metrics, another for instructing participants, and a third for administering questionnaires and conducting interviews. 

To mitigate order effects, the study employed a counterbalancing design. Participants were randomly assigned to begin with one of two experimental conditions, resulting in half starting with GO and the other half with VG. In each condition, participants performed identical tasks. In the GO condition, participants used only hand gesture interaction to complete the task. In the VG condition, participants used voice commands in the first task stage to move the robot arm from the origin to the starting zone by articulating \textit{"Blueberry, move to [a certain color of the cubes]"}, rather than manually controlling it. Next, both the virtual and physical robots moved in sync, with the physical robot moving itself above a selected cube. Participants were instructed to begin with any cube of their choice. Then, they proceeded with the hand gesture interaction onwards. After completing their initial condition, each participant switched to the other condition to allow for a within-subject comparison. The overall study workflow is illustrated in Fig.~\ref{fig:us_process}.

\begin{table*}[!t]
    \centering
    \caption{Self-developed measurement metrics in the user study.}
    \label{tab:metrics}
    \begingroup
    \setlength{\tabcolsep}{2pt}
    \renewcommand{\arraystretch}{0.85}
    \fontsize{6}{7.2}\selectfont  
    \begin{tabular}{p{3cm} p{10.5cm} p{1cm}}
        \toprule
        \rowcolor{gray!30} \textbf{Metric Category} & \textbf{Definition/Explanation} & \textbf{Range} \\
        \midrule
        \rowcolor{gray!15} \multicolumn{3}{l}{\textbf{Performance Metrics}} \\
        \midrule
        \textbf{Number of Cubes} & Number of cubes successfully placed in the corresponding colored cells in the target zone. & - \\
        \textbf{Unsuccessful Attempts} & The robotic gripper attempted to grasp but failed; or the cube was not placed fully inside the designated colored cell. & - \\
        \textbf{TCT} & Total time taken to complete the task within the time limit. The maximum is 300s & - \\
        \textbf{Interruptions} & The number of times the system stopped due to user misoperations (i.e., too fast hand movements). & - \\
        \midrule
        \rowcolor{gray!15} \multicolumn{3}{l}{\textbf{User Experience \& Usability Metrics}} \\
        \midrule
        \multicolumn{3}{l}{\textit{\textbf{Customized Metrics}}} \\
        \hline
        \textbf{Perceived Ease of Use} & \textit{"How easy do you think it is to use the system?"} & [1, 7] \\
        \textbf{Perceived Effectiveness} & \textit{"I think using this system is effective for real tasks."} & [1, 7] \\
        \textbf{Perceived Improved Safety} & \textit{"I think this system would enhance safety compared to direct robot control."} & [1, 7] \\
        \textbf{Perceived Intent to Use} & \textit{"I would like to use this system for robot control."} & [1, 7] \\
        \textbf{Perceived Comfort} & \textit{"I find the gesture/voice interaction modalities comfortable."} & [1, 7] \\
        \textbf{Satisfaction} & \textit{"In general, I am satisfied with using this system."} & [1, 7] \\
        \bottomrule
    \end{tabular}
    \endgroup
\end{table*}

\subsection{Measurements}
\label{measure}
To evaluate the impact of the multimodal interaction, we employed a combination of performance metrics and UX and usability metrics. Performance was measured using four key indicators: number of cubes, unsuccessful attempts, task completion time (TCT, the maximum is 300s), and number of interruptions, providing insights into task efficiency, accuracy, and user control. For UX and usability assessment, we selected UEQ-S (pragmatic quality and hedonic quality, in [-3, +3 Likert scale], with each consisting of four metrics) \cite{schrepp2017design,pick2016design,lin2024usability}, a shortened version of UEQ, to balance comprehensive evaluation with experiment duration constraints. Additionally, NASA TLX was utilized to measure the perceived workload in 7-point Likert scale \cite{shayesteh2021investigating,polvi2017handheld}. To further explore usability aspects, we included six customized questions inspired by previous works \cite{zhao2024glanxr,chang2024perceived,lin2024usability}, focusing on system usability and user perceptions, which were also measured using 7-point Likert scale for unifying the process as well as better quantifying the evaluation. The complete hierarchy and definitions of self-developed metrics are detailed in Tab.~\ref{tab:metrics}. We conducted Shapiro–Wilk normality tests for each metric and then applied the corresponding statistical analysis.

\section{Results}
\subsection{Quantitative Analysis}
\subsubsection{Performance Metrics}
\begin{figure*}[!t]
  \centering
  \captionsetup[sub]{justification=centering}

  \begin{subfigure}[t]{0.48\linewidth}
    \centering
    \includegraphics[width=.8\linewidth]{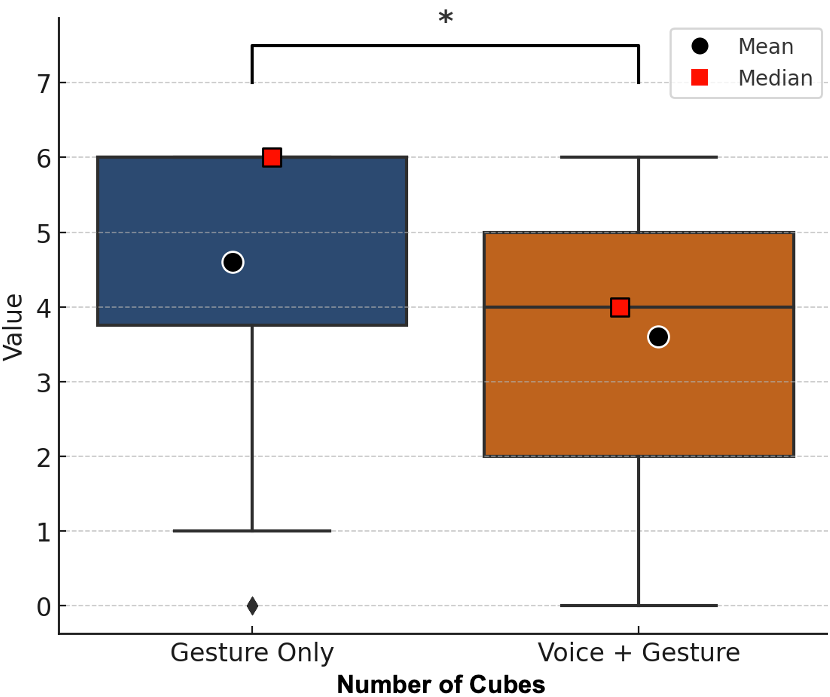}
    \caption{No. of cubes.}
    \label{fig:no_cubes}
  \end{subfigure}\hfill
  \begin{subfigure}[t]{0.48\linewidth}
    \centering
    \includegraphics[width=.8\linewidth]{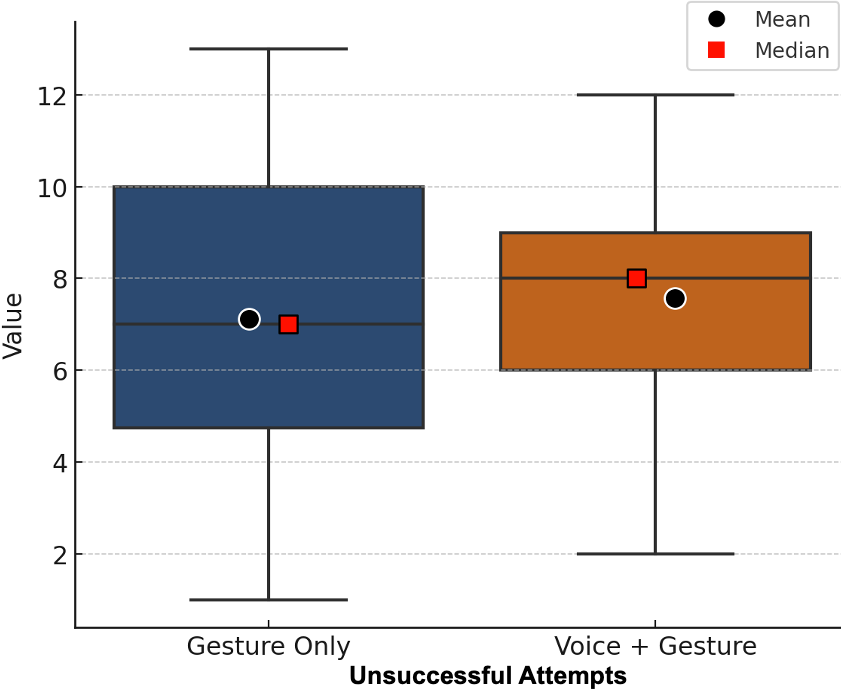}
    \caption{Unsuccessful attempts.}
    \label{fig:unsuccessful}
  \end{subfigure}

  \medskip

  \begin{subfigure}[t]{0.48\linewidth}
    \centering
    \includegraphics[width=.8\linewidth]{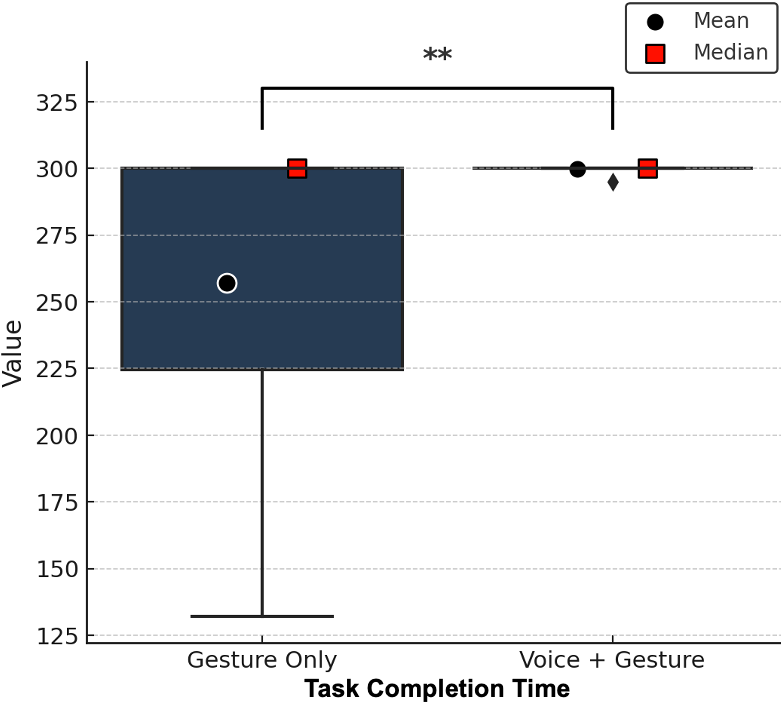}
    \caption{TCT.}
    \label{fig:task_time}
  \end{subfigure}\hfill
  \begin{subfigure}[t]{0.48\linewidth}
    \centering
    \includegraphics[width=.8\linewidth]{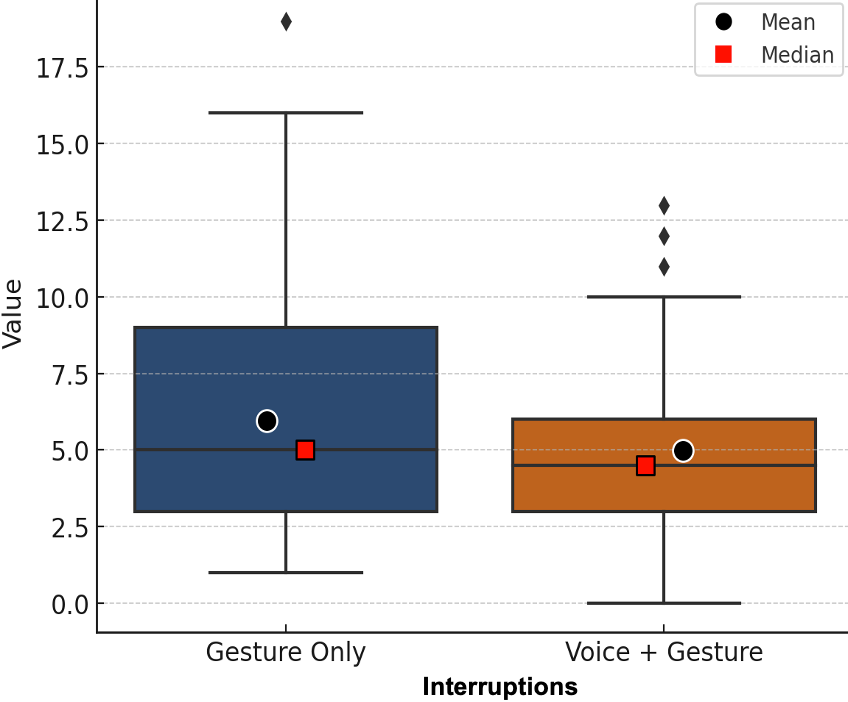}
    \caption{Interruptions.}
    \label{fig:interrup}
  \end{subfigure}

  \caption{Comparison of performance metrics in the study.}
  \label{fig:performance}
\end{figure*}

\mbox{}\\[0.5\baselineskip]
\textbf{\textit{Number of Cubes.}} Normality was confirmed ($p > 0.05$). This metric was significantly higher in the GO condition ($M = 4.60, SD = 2.19$) compared to VG by paired-t test ($M = 3.60, SD = 1.79, p < 0.05$, $Cohen’s$ $d = 0.60$). The result indicates that incorporating voice commands may have introduced additional cognitive load or inefficiencies, leading to a lower number of successfully moved cubes. The observed variability in performance as shown in Fig.~\ref{fig:no_cubes} suggests differential adaptation to multimodal interaction among participants.

\textbf{\textit{Unsuccessful Attempts.}} Normality was obtained ($p > 0.05$) and paired t-test revels no significant difference was found between the two conditions ($Cohen’s$ $d = 0.12$), as shown in Fig.~\ref{fig:unsuccessful},. The mean number of failed attempts was slightly higher in VG ($M = 7.57, SD = 2.41$) than in GO ($M = 7.13, SD = 3.06$), but this difference was not statistically significant. This implies that error rates were likely influenced by factors such as individual skill levels or task complexity, rather than the interaction modality.

\textbf{\textit{TCT.}} Wilcoxon signed-rank test was used since normality was not shown in TCT. Given the 300-second time limit, most participants in VG reached the maximal time, whereas the GO group had a lower TCT ($M = 261s$), achieving successful completion more frequently ($p < 0.01$, $r = 0.19$) (Fig.~\ref{fig:task_time}). This finding indicates that gesture-only interaction facilitated more efficient task execution, likely due to fewer errors and quicker responses compared to the multimodal VG condition.

\textbf{\textit{Interruptions.}} This metric showed normality ($p > 0.05$) while the paired t-test indicated that interruptions occurred slightly more frequently in GO ($M = 5.95, SD = 4.30$) than in VG ($M = 4.98, SD = 3.03$) (Fig.~\ref{fig:interrup}). However, this difference was not statistically significant ($Cohen’s$ $d = 0.22$). The presence of outliers in GO suggests that some users encountered more issues without voice commands, hinting at a possible but inconsistent advantage of multimodal input in reducing interaction errors.

In summary, GO emerged as the more efficient interaction mode, with participants moving more cubes and completing tasks more quickly. While VG showed a slight, non-significant reduction in unsuccessful attempts and interruptions, the added voice modality did not demonstrably enhance overall task performance and may have contributed to increased complexity.

\subsubsection{UX and Usability Metrics}
\begin{figure*}[!t]
\centering
\captionsetup[sub]{justification=centering}

\begin{subfigure}[t]{0.48\linewidth}
  \centering
  \includegraphics[width=\linewidth]{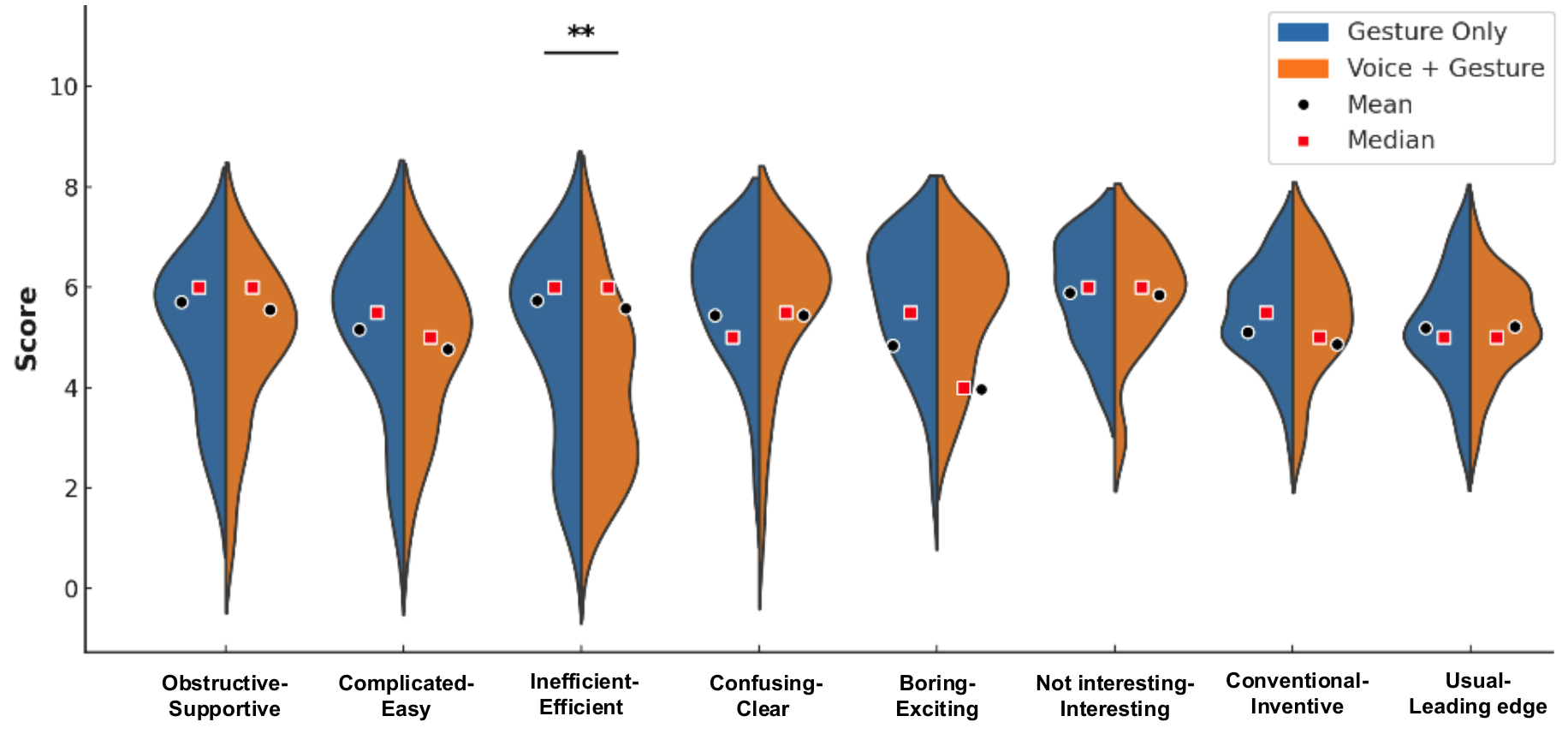}
  \caption{UEQ-S.}
  \label{fig:ueq}
\end{subfigure}\hfill
\begin{subfigure}[t]{0.48\linewidth}
  \centering
  \includegraphics[width=\linewidth]{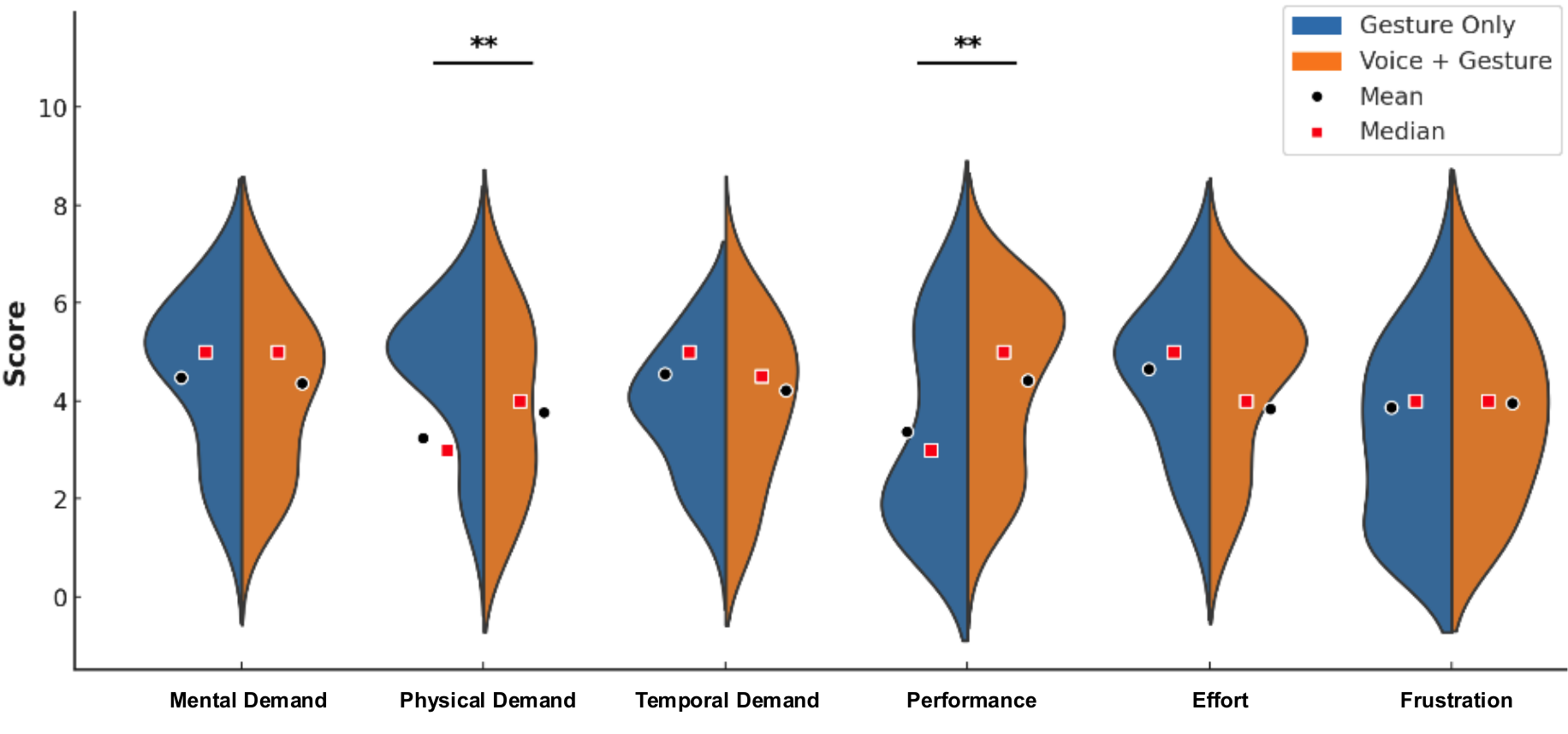}
  \caption{NASA TLX.}
  \label{fig:tlx}
\end{subfigure}

\medskip

\begin{subfigure}[t]{0.48\linewidth}
  \centering
  \includegraphics[width=\linewidth]{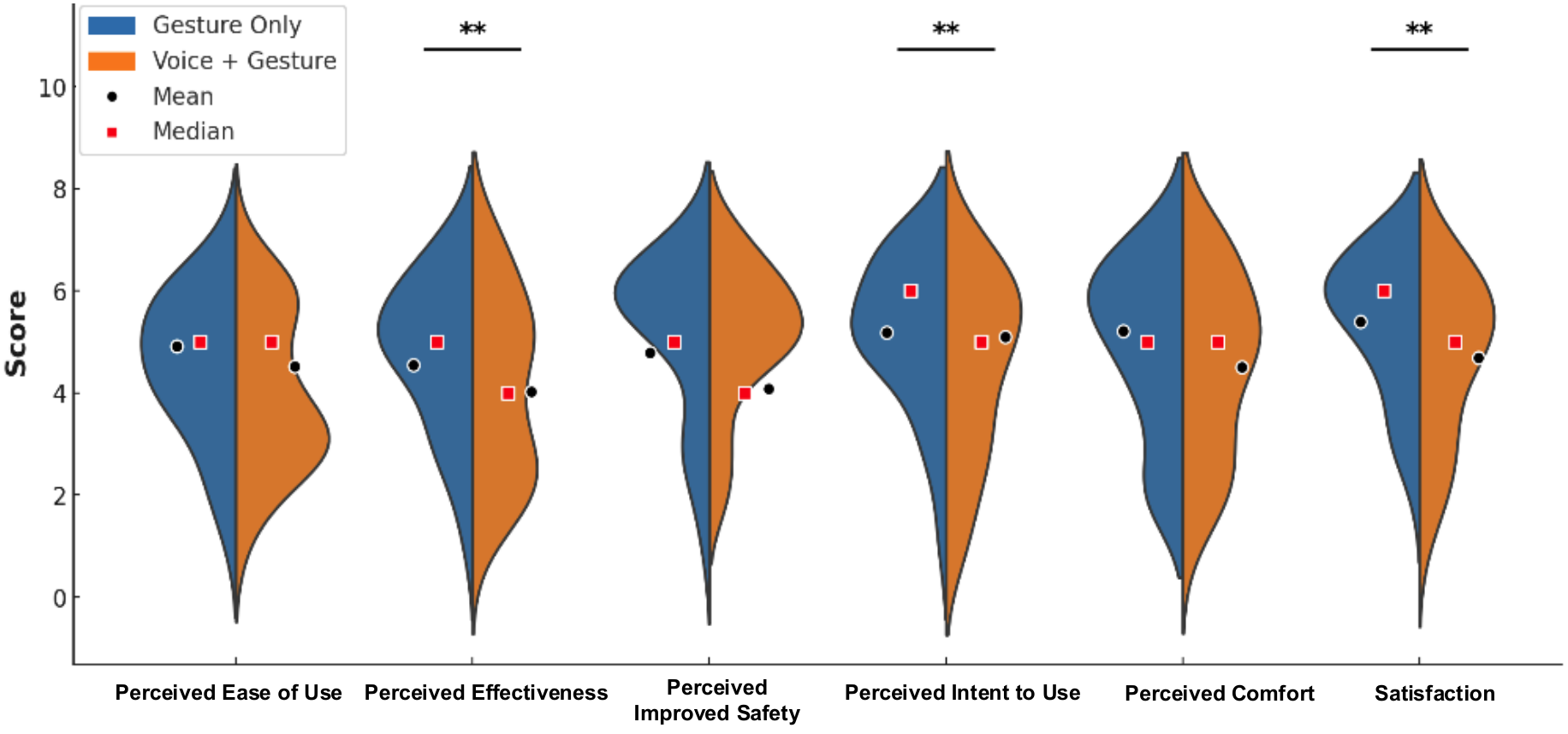}
  \caption{Customized.}
  \label{fig:cus}
\end{subfigure}\hfill
\begin{subfigure}[t]{0.48\linewidth}
  \centering
  \includegraphics[width=\linewidth]{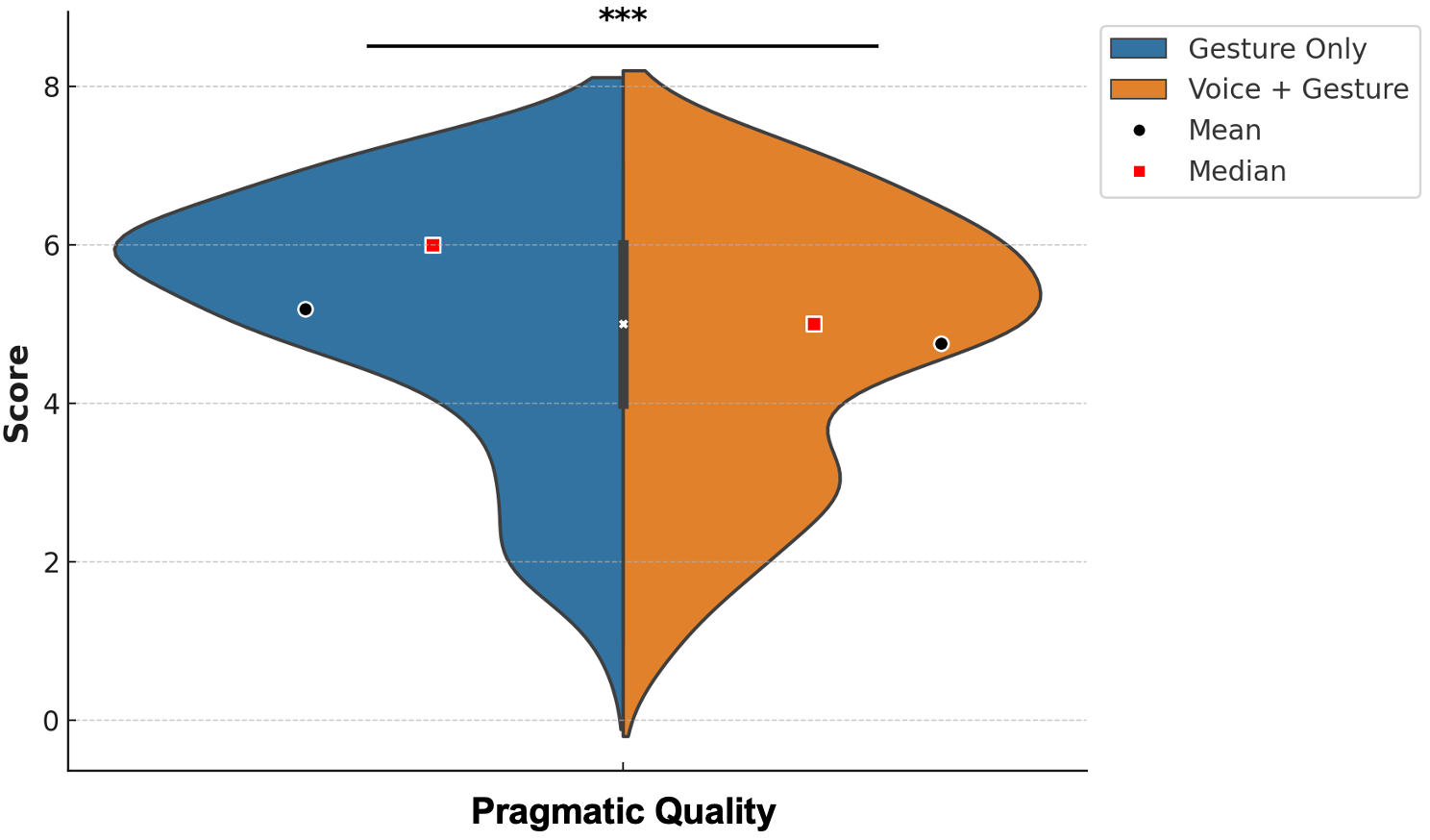}
  \caption{UEQ-S: pragmatic quality.}
  \label{fig:ueq_prag}
\end{subfigure}

\medskip

\begin{subfigure}[t]{0.48\linewidth}
  \centering
  \includegraphics[width=\linewidth]{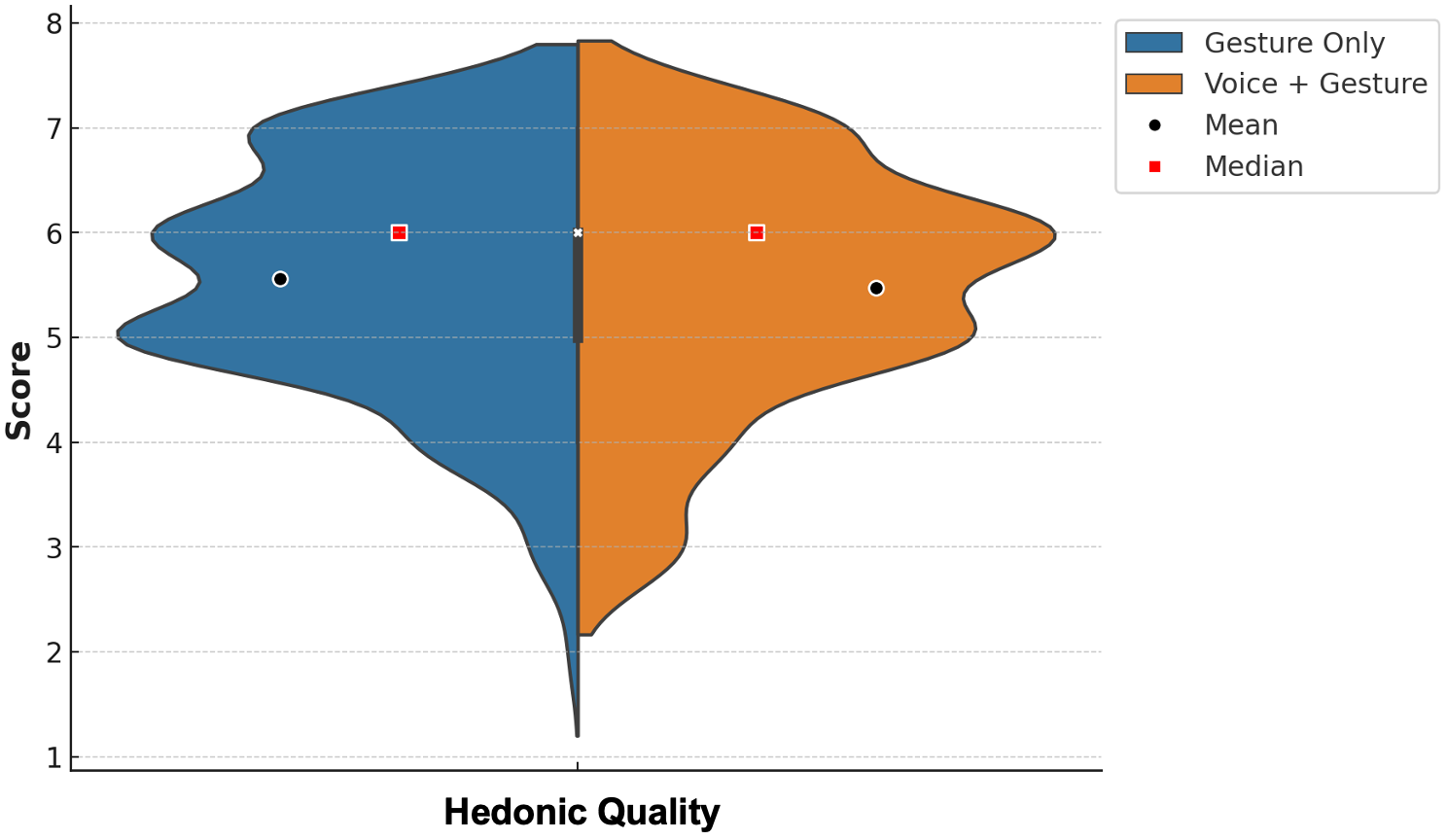}
  \caption{UEQ-S: hedonic quality.}
  \label{fig:ueq_hedo}
\end{subfigure}\hfill
\begin{subfigure}[t]{0.48\linewidth}
  \centering
  \includegraphics[width=\linewidth]{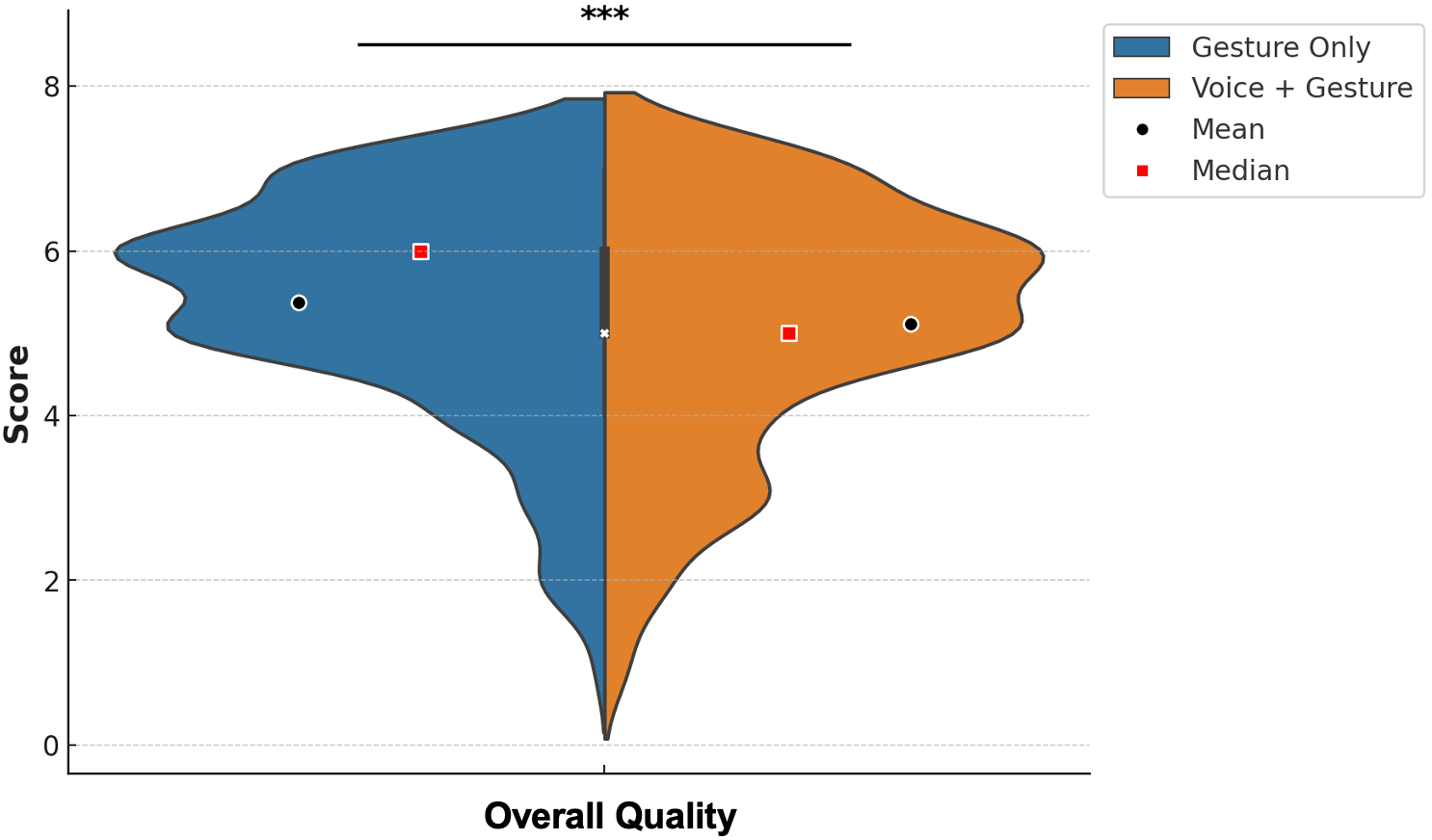}
  \caption{UEQ-S: overall quality.}
  \label{fig:ueq_over}
\end{subfigure}

\caption{Comparison of UX and usability metrics, including UEQ-S, NASA TLX, customized metrics, and UEQ-S subscales.}
\label{fig:ux_combined}
\end{figure*}

\mbox{}\\[0.5\baselineskip]
\textbf{\textit{UEQ-S.}} Following the normality check (Section ~\ref{measure}), an item-wise analysis of UEQ-S revealed that GO tended to exceed the values across most variables compared to VG (Fig.~\ref{fig:ueq}). Despite considerable overlap in distributions, certain items -- such as Boring-Exciting -- showed clearer separation. Notably, the Inefficient–Efficient scale favored GO significantly (run by paired-t test; $p < 0.01$, $Cohen’s$ $d = 0.23$).

For conventional analysis, pragmatic quality revealed a highly significant difference by Wilcoxon signed-rank test ($p < 0.001$, $r = 0.35$), with GO ($M = 5.19, SD = 1.52$) rated higher than VG ($M = 4.76, SD = 1.64$) (Fig.~\ref{fig:ueq_prag}). This suggests that gesture-only interaction was perceived as more efficient, predictable, and user-friendly. Gestures, being direct and immediate, likely contributed to this perception, whereas voice commands may have introduced recognition delays or increased cognitive load. Paired t-test showed hedonic quality differed insignificantly between conditions ($Cohen’s$ $d = 0.10$), with GO ($M = 5.56, SD = 1.10$) and VG ($M = 5.47, SD = 1.14$), which were rated similarly (Fig.~\ref{fig:ueq_hedo}). This suggests both modalities were comparably engaging and enjoyable. The multimodal novelty of VG did not significantly enhance  user enjoyment to some extent, implying that gesture-based input alone can provide intuitiveness for such experiences. Paired t-test also showed the overall quality significantly favored GO ($M = 5.38, SD = 1.34$) over VG ($M = 5.12, SD = 1.46$) ($p < 0.001$, $Cohen’s$ $d = 0.71$) (Fig.~\ref{fig:ueq_over}). This reflects the strong impact of GO’s pragmatic advantages, as hedonic differences were negligible. The results suggest that while voice input may offer contextual benefits, it does not inherently improve UX when combined with gestures. On the contrary, it may introduce complexity and inefficiencies that detract from perceived usability.

\textbf{\textit{NASA TLX.}} All scales underwent Wilcoxon signed-rank tests since normality was not obtained in any scale. As Fig.~\ref{fig:tlx} shows, the VG condition recorded higher scores across most workload categories, contrary to expectations that multimodality would trigger lower scores indicating reduced workload. Notably, VG exhibited more compressed distributions in certain dimensions, pointing to more consistent user perceptions. Significant differences were found in physical demand by paired t-test ($p < 0.005$, $r = 0.51$), with GO perceived as more physically demanding. This suggests that integrating voice commands alleviates some physical effort. For Perceived Performance, VG was rated significantly higher than GO ($p < 0.005$, $r = 0.42$), implying that participants felt more successful in task execution when only voice interaction was available. Although mental demand showed no significant difference ($r = 0.26$) indicates a slight increase in mental demand for GO. No significances were observed for Temporal Demand, Effort, and Frustration, suggesting similar perceptions of time pressure, exertion, and frustration across both conditions.

\textbf{\textit{Customized Metrics.}} Compared to standard instruments, customized metrics displayed greater variability (Fig.~\ref{fig:cus}). Wilcoxon signed-rank tests were conducted due to lack of normality in each metric. Some GO metrics had higher medians but wider spread, while VG values tended to be lower and more consistent, indicating possible usability concerns. Significant differences were observed in perceived effectiveness ($p < 0.005$, $r = 0.33$), perceived intent to use ($p < 0.005$, $r = 0.42$), and satisfaction ($p < 0.005$, $r = 0.51$), with GO outperforming VG in all three. This suggests that participants found gesture-only interaction more effective, easier to use, and more satisfying overall. However, the broad variance indicates individual user preferences played a substantial role, especially in other customized metrics not reaching significant differences.

\subsection{Qualitative Analysis}

\begin{figure*}[!t]
    \centering
    \includegraphics[width=\linewidth]{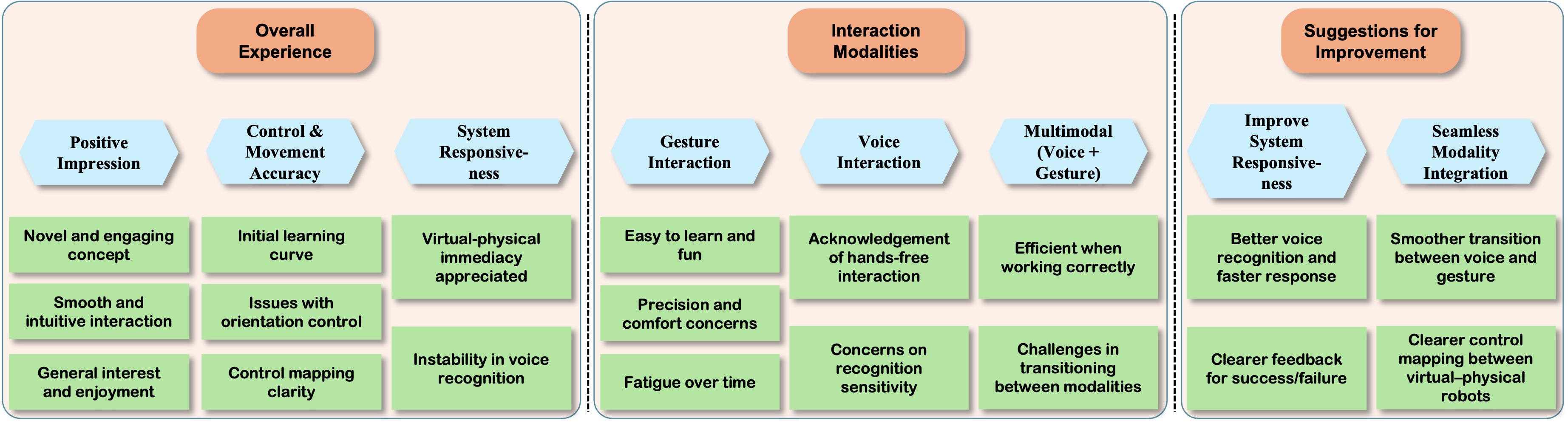}
    \caption{Themes and details from qualitative results.}
    \label{fig:qua}
\end{figure*}

We now present the qualitative findings derived from the post-study interviews to gain deeper insight into participants’ experiences and perceptions, obtained from a thematic analysis. One author led open coding using an affinity-diagram approach on a shared Miro board. Two
additional authors independently reviewed the codes and proposed thematic groupings. ChatGPT-4o was used as a supportive labeling tool for organizing and summarizing candidate themes, but was not used to generate or interpret themes autonomously; all interpretive decisions
were made by the human authors. Final themes were validated through several iterative participatory discussions among all authors to reach consensus. While we did not compute a formal inter-rater reliability, the iterative multi-author process and the explicit documentation of themes in Fig.~\ref{fig:qua} provide the methodological transparency.

\subsubsection{Overall Experience}\mbox{}\\[0.5\baselineskip]
\textbf{Positive impression on intuitiveness \& novelty.}
Participants found the system novel and engaging, appreciating its innovative interaction methods. One participant stated, \textit{“I feel like it’s a very novel experience. You can interact with virtual stuff, like the robot model, as well as controlling the real robot. It’s very interesting.”} Another noted, \textit{“the whole experience is quite nice, everything is very smooth,”} underlining the potential for intuitive interactions. Users favoring GO interaction specifically highlighted ease of use, with one remarking, \textit{“I prefer the GO condition compared to the VG condition,”} citing fewer errors. Another shared, \textit{“It’s very intuitive to just do the pinching gestures and move your hands.”} A user also mentioned, \textit{“The task seems to be interesting and refreshing to me,”} reinforcing the engaging nature of the system.

\textbf{Challenges with control and movement accuracy.}
While participants appreciated the system’s capabilities, a common theme was that achieving precise manipulation required effort. One participant remarked, \textit{“It’s not easy to control the robotic arm’s movement accurately,”} suggesting that clearer axis coordination could further assist users. Another stated, \textit{“The task seems to be interesting, but in reality, it was a bit difficult to fully control the robot rotation angles.”} Users occasionally noted inconsistencies in responsiveness, with one commenting, \textit{“Sometimes it was unclear how fast you could move before going into the interruption. The feeling of going into slow movements can make it less efficient.”} Sensitivity was also flagged: \textit{“I think it’s too sensitive and it captures some instability of my hands,”} indicating an opportunity to refine control mapping and provide clearer visual or tactile cues.

\textbf{System responsiveness and feedback.} Users generally reported that the system was responsive to interaction inputs, with the virtual counterpart reacting immediately upon correct gesture initiation, followed promptly by the physical robot, which was positively noted. At the same time, occasional instability – particularly in voice command recognition – was observed, highlighting opportunities to strengthen feedback mechanisms and overall system robustness.

\subsubsection{Interaction Modalities}\mbox{}\\[0.5\baselineskip]
\textbf{Gesture interaction acknowledged with praise.} 
GO was generally well-received and described as intuitive and enjoyable, with users favoring its straightforwardness and immediacy. Participants found it easier to understand and use gestures for basic control tasks, with one commenting, \textit{“It’s very straightforward and intuitive to do the pinch.”} However, precision and ergonomic comfort presented challenges, with concerns about sensitivity and fatigue during prolonged use. Some participants noted gestures becoming less reliable over time, with one stating, \textit{“The gesture control starts getting into weird configurations where it becomes harder to maneuver.”}

\textbf{Improving voice interaction for consistency.} 
Voice interaction was generally appreciated for its convenience and hands-free control. However, some participants noted variability in recognition accuracy, particularly with diverse accents and speech patterns. While this occasionally affected usability, several users reported that the experience improved with brief familiarization. One participant remarked, \textit{“The voice commands were not robust enough against accents, so that part was demotivating,”} while another noted that initial difficulties gave way to smoother interactions over time. The discrete nature of voice commands sometimes disrupted the flow when switching between modalities, suggesting opportunities to enhance continuity and user feedback for a more seamless experience.

\textbf{Multimodal interaction fluctuated among evaluation.} 
When combining voice and gestures, some articipants experienced increased cognitive load and complexity. Some appreciated the multimodal interaction’s efficiency, especially when both modalities functioned seamlessly. Some participants found switching between voice and gesture inputs was sometimes unnatural, indicating a need for better integration and smoother transitions. One participant explicitly mentioned, \textit{“It was a bit confusing because you have to let it go while you speak and then go back into grabbing cubes.”}

\begin{figure*}[!t]
\centering
\captionsetup[sub]{justification=centering}

\begin{subfigure}[t]{0.48\linewidth}
  \centering
  \includegraphics[width=.9\linewidth]{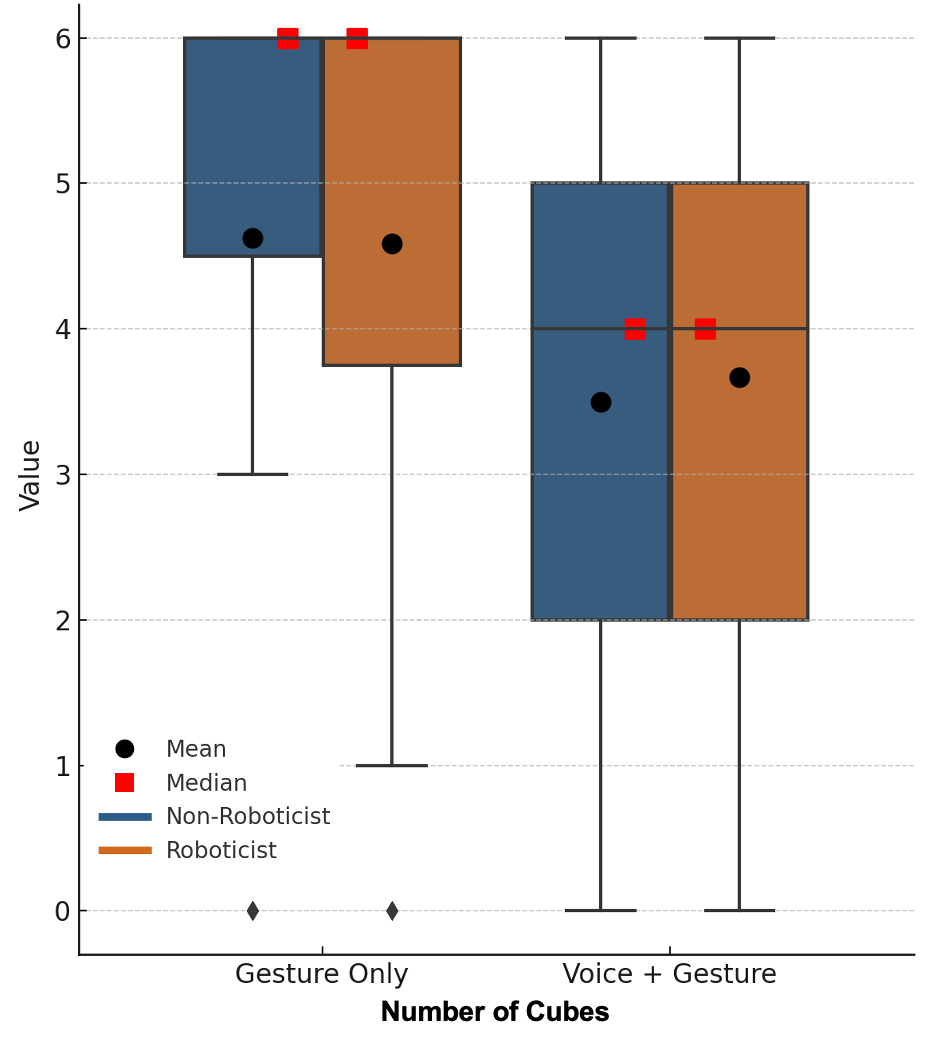}
  \caption{No. of cubes.}
  \label{fig:ro_no_no_cubes}
\end{subfigure}\hfill
\begin{subfigure}[t]{0.48\linewidth}
  \centering
  \includegraphics[width=.9\linewidth]{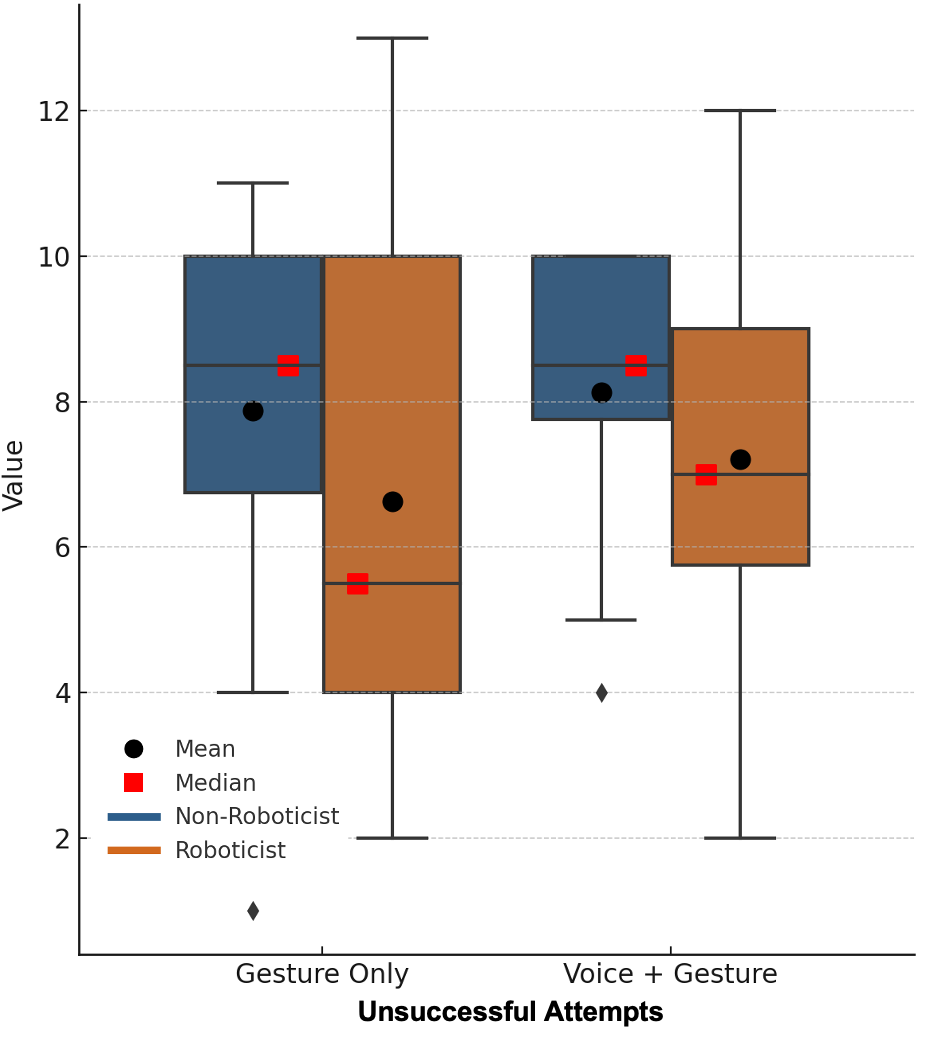}
  \caption{Unsuccessful attempts.}
  \label{fig:ro_no_unsuccessful}
\end{subfigure}

\medskip

\begin{subfigure}[t]{0.48\linewidth}
  \centering
  \includegraphics[width=.9\linewidth]{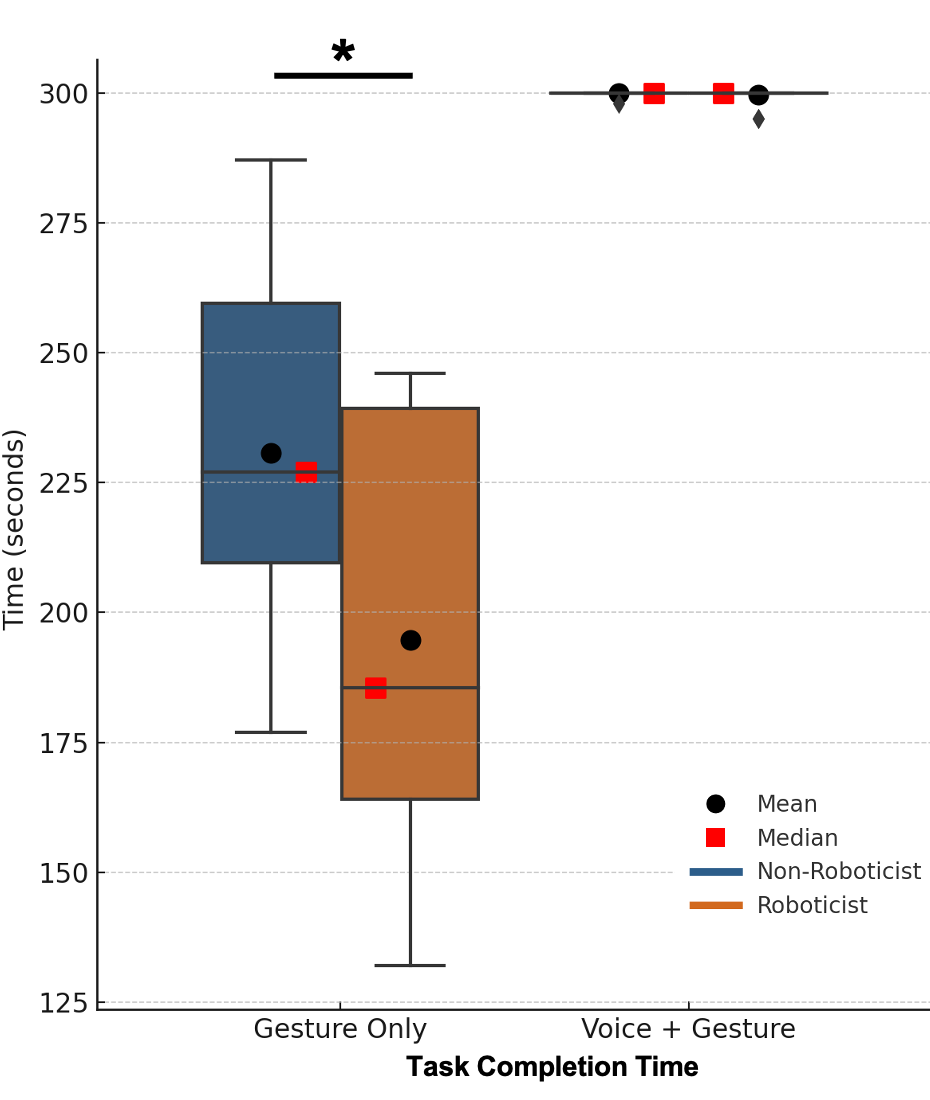}
  \caption{TCT.}
  \label{fig:ro_no_task_time}
\end{subfigure}\hfill
\begin{subfigure}[t]{0.48\linewidth}
  \centering
  \includegraphics[width=.9\linewidth]{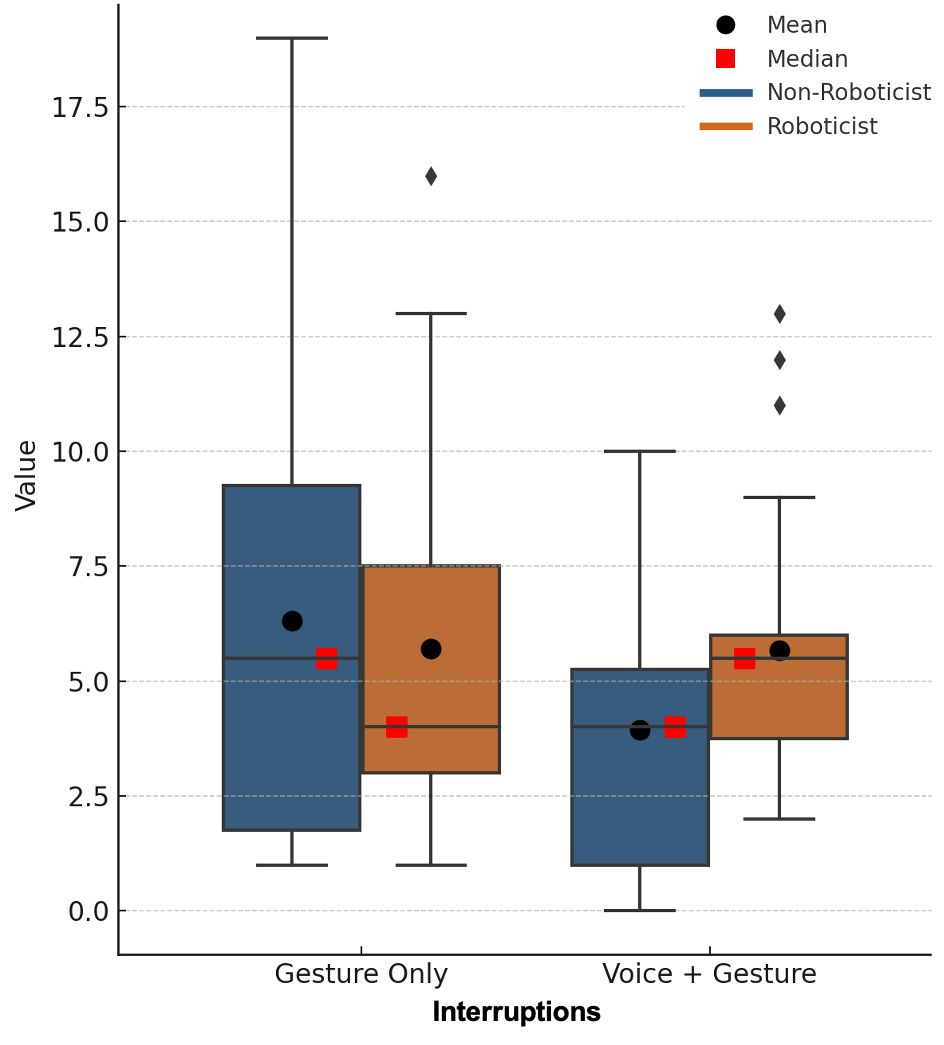}
  \caption{Interruptions.}
  \label{fig:ro_no_interrup}
\end{subfigure}

\caption{Comparative results of performance between roboticists and non-roboticists.}
\label{fig:robo_non_robot_per}
\end{figure*}

\begin{figure*}[!t]
\centering
\captionsetup[sub]{justification=centering}

\begin{subfigure}[t]{0.8\textwidth}
  \centering
  \includegraphics[width=\textwidth]{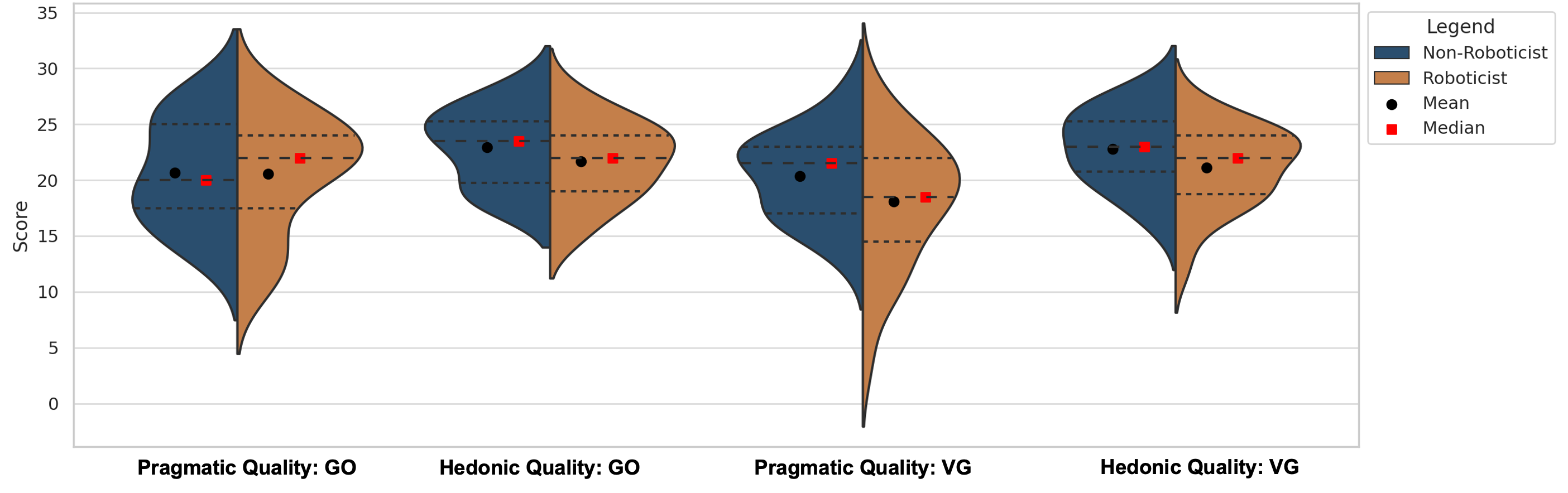}
  \caption{UEQ-S.}
  \label{fig:ro_no_ueq}
\end{subfigure}

\medskip

\begin{subfigure}[t]{0.8\textwidth}
  \centering
  \includegraphics[width=\textwidth]{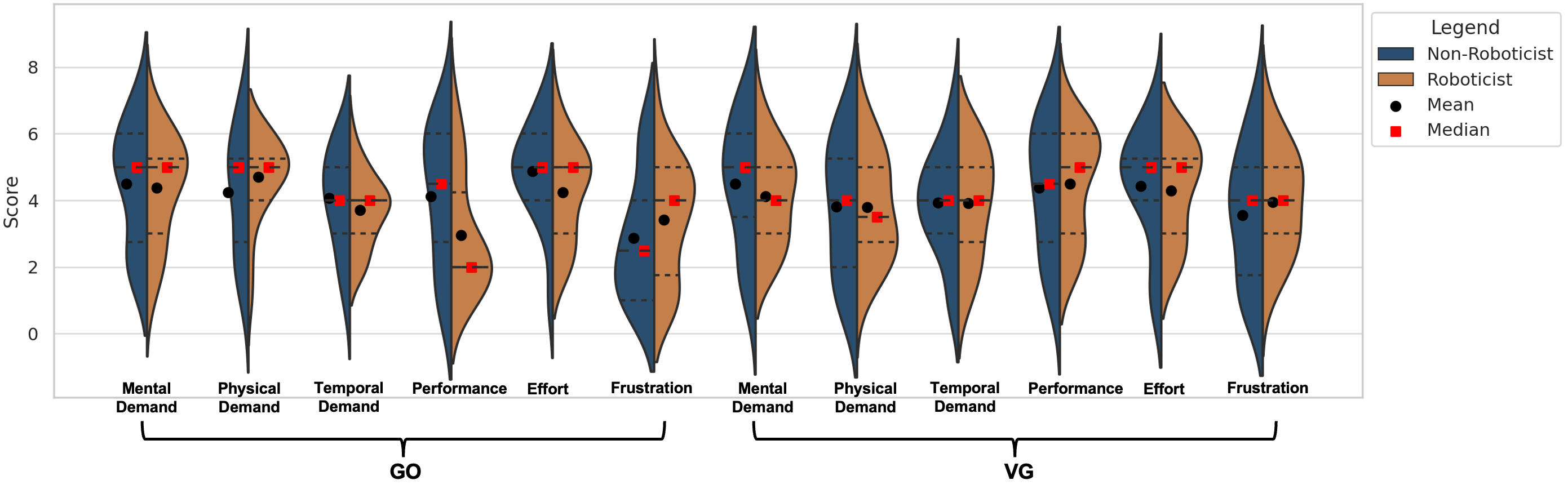}
  \caption{NASA TLX.}
  \label{fig:ro_no_tlx}
\end{subfigure}

\medskip

\begin{subfigure}[t]{0.8\textwidth}
  \centering
  \includegraphics[width=\textwidth]{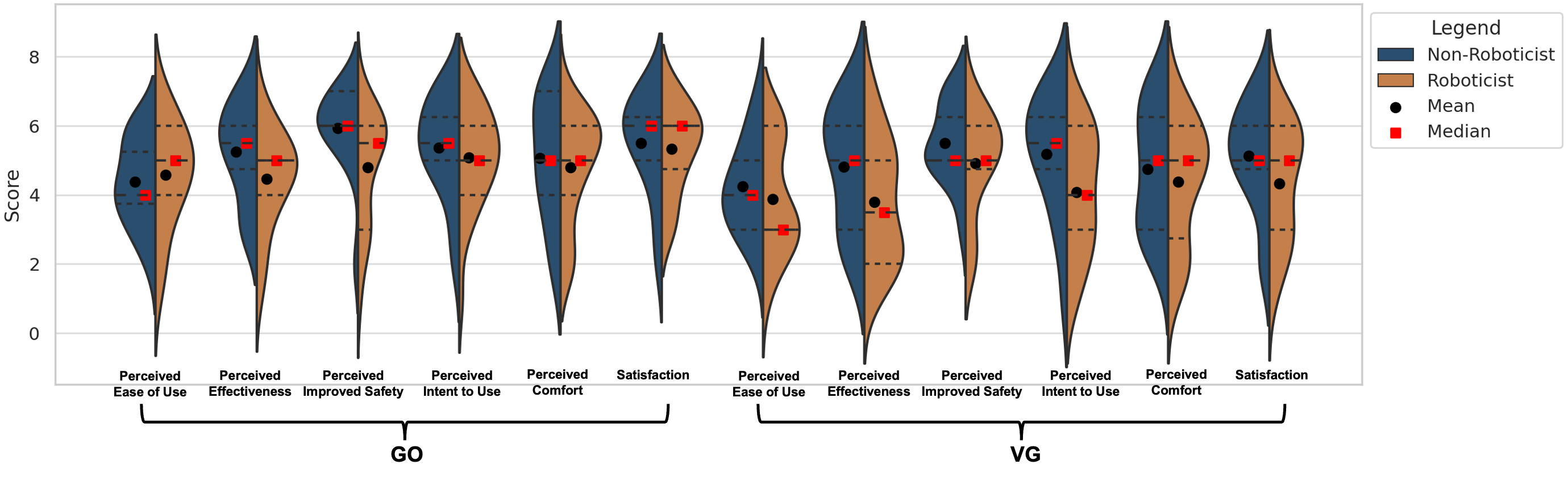}
  \caption{Customized metrics.}
  \label{fig:ro_no_cus}
\end{subfigure}

\caption{Comparison of UX and usability metrics between roboticists and non-roboticists.}
\label{fig:ux_metrics}
\end{figure*}

\subsubsection{Suggestions for Improvement}\mbox{}\\[0.5\baselineskip]
\textbf{Enhancing system stability and responsiveness.}
Participants expressed a desire for improved system responsiveness and reliability which included refining voice recognition and reducing system latency, with one participant recommending \textit{“better voice recognition, as well as a faster response time,”} and another noting the need for the system to be \textit{“more responsive at detecting when you speak to it.”} Gesture interaction also received feedback, with users highlighting the importance of real-time feedback to support confidence and reduce the need to repeat actions. As one participant suggested, \textit{“instant feedback, so I know if I’m doing it right instead of having to restart every time it fails.”}

\textbf{Smoother modality transition and more fluid robotic control.}
One participant noted challenges in switching between voice and gesture inputs, suggesting that transitions should feel more natural and uninterrupted. Additionally, several users highlighted the need for improved control mapping between the virtual and physical robots. This could be addressed through immediate and clear visual or tactile cues, potentially enhancing overall usability, interaction efficiency, and user awareness.

\subsection{Analysis Between Roboticist vs. Non-Roboticist}
In addition, we explored the differences between roboticists and non-roboticists from our participant pool. Statistical analysis was conducted after the normality tests.

\subsubsection{Performance}\mbox{}\\[0.5\baselineskip]
The results indicate that roboticists significantly outperformed non-roboticists in TCT when using merely gesture interaction, completing the task in 194.64 seconds compared to 230.73 seconds for non-roboticists (independent t-test; $p < 0.05$ with a large effect size $Cohen's$ $d = 0.926$) (Fig.~\ref{fig:ro_no_no_cubes}). However, when voice command was introduced, this difference disappeared, with both groups averaging nearly 300 seconds with no significance. This suggests that GO required a level of expertise that gave roboticists an advantage, while VG effectively equalized performance, making the system more accessible to all users. Despite this, no significant differences were found in other three metrics between the two groups. Similarly, roboticists surpassed non-roboticists obviously in TCT within GO, while both had approximate TCT in VG (Fig.~\ref{fig:ro_no_task_time}), indicating that the task completion was not necessarily dependent on prior robotics experience in multimodal interaction. Regarding unsuccessful attempts, roboticists had an obvious lower mean value in GO, especially in VG, most roboticist had less unsuccessful attempts with the help of voice command interaction (Fig.~\ref{fig:ro_no_unsuccessful}). An interesting trend was observed in the number of interruptions in the VG condition. Roboticists ($M = 5.67$) exhibited more interruptions than non-roboticists ($M = 3.94$) (Fig.~\ref{fig:ro_no_interrup}), though this result was not significant via Mann–Whitney U test ($r = 0.24$). This could suggest that roboticists found the voice commands less efficient or more intrusive, leading to disruptions in their task flow. Non-roboticists, on the other hand, may have benefited from voice assistance, which possibly helped reduce execution errors, leading to fewer system interruptions. This suggests that while roboticists were faster, the overall execution errors and precision challenges were shared across all users, likely influenced more by the interface itself than by individual expertise.


\subsubsection{UX and Usability}\mbox{}\\[0.5\baselineskip]
For \textbf{\textit{UEQ-S}}, both pragmatic and hedonic quality scores showed no significance between the two groups across the two conditions after performing independent t-tests (Fig.~\ref{fig:ro_no_ueq}). Roboticists and non-roboticists performed similarly in pragmatic quality under the GO condition, suggesting that the system’s usability was accessible regardless of prior expertise. Notably, non-roboticists reported higher enjoyment and engagement in hedonic quality compared to roboticists. VG consistently yielded slightly higher mean and median values for both quality dimensions than GO, especially for non-roboticists. This indicates that the inclusion of voice interaction may have provided a more intuitive and accessible interface for users unfamiliar with robotic technologies.

\textbf{\textit{NASA TLX}} revealed more nuanced but functionally meaningful differences. Although no significance emerged (normality gained, independent t-tests executed), mental demand and effort scores were visibly higher for non-roboticists across both conditions (Fig.~\ref{fig:ro_no_tlx}). This implies that controlling the robot via gesture and voice may have posed greater cognitive challenges for less experienced users. Interestingly, with the addition of voice commands, non-roboticists showed smaller differences from roboticists in temporal demand, performance, and frustration levels, effectively narrowing the workload gap. Notably, frustration levels were marginally higher among roboticists in both conditions, possibly due to their more critical evaluation standards or elevated expectations for system responsiveness and precision based on prior experience.

The most prominent distinctions between the two groups appeared in the \textbf{\textit{Customized Metrics}}. While statistical significance was not achieved by independent t-tests, several sub-metrics revealed clear group differences. Under the GO condition, non-roboticists rated ease of use lower than roboticists, as reflected in both the median values and distribution ranges. However, for all other sub-metrics, non-roboticists gave higher ratings. In the VG condition, non-roboticists outperformed roboticists across all sub-metrics, including a notable improvement in ease of use. This suggests that voice command integration significantly enhanced the subjective experience for novice users, making the system feel more intuitive, controllable, and engaging. On the other hand, the relatively lower ratings from roboticists may reflect their more demanding performance expectations and the importance they place on contextual efficiency when evaluating new systems. 

\subsubsection{Summary of Qualitative Feedback}\mbox{}\\[0.5\baselineskip]
\textbf{Non-roboticists generally responded positively to the GO interaction, finding it more natural and intuitive, with affirmation on VG interaction.} One participant remarked, \textit{“I like the gesture one since the implementation is simple and effective,”} while another described the interaction as \textit{“great”}. Even there was some difficulties with voice recognition—several users noted that the system had trouble distinguishing commands from environmental sounds, or that the voice interface felt unpredictable, they praised this modality due to its convenience. For instance, one participant stated, \textit{“For the voice command, it may cause some errors because it might recognize surrounding sounds, but it's so convenient to instruct a robot with a simple voice input besides using gestures”} and another recommended, \textit{“It didn't recognize my pronunciation at first try, but it eventually worked and executed the command correctly, which helped processing the task.”} These comments suggest that novice user appreciated the multimodality due to it offers another option during interaction.

\textbf{In contrast, roboticists engaged with the system more critically, prioritizing precision, robustness, and ergonomic sustainability.} While some appreciated the multimodal aspect, they were quick to identify system inefficiencies, particularly related to hand gesture fatigue and grasp control mechanics. One experienced user mentioned, \textit{“My arm is tired and it is a bit hard to control the grabbing movement,”} and another pointed out that \textit{“The main problem is with the hand gripping—it doesn’t always grab properly.”} Despite recognizing the value of multimodal input, roboticists seemed to expect greater technical reliability and feedback mechanisms, including improvements such as end-effector orientation cues or grab confirmation signals: \textit{“It would be better if there’s a feedback when the robot grabs,”} and \textit{“Visual feedback from the voice (text on the screen) would help.”} A few also discussed system crashes and unpredictable behavior under VG conditions, reflecting a desire for greater system stability and feedback transparency.

In summary, the data illustrates a cognitive and experiential divide between the two groups. Non-roboticists favor simplicity, clarity, and responsiveness, often forgiving of technical imperfections if the interface remains accessible. Roboticists, conversely, demand technical precision, stability, and ergonomic viability, expressing frustration when those expectations are unmet. These divergent user needs suggest that future versions of the AR 'puppeteer' system should adopt adaptive interface strategies, offering more guided, simplified interactions for novices, and more fine-tuned, responsive controls for expert users. Additionally, both groups’ repeated emphasis on improving the voice modality signals its current status as a bottleneck for interaction fluidity and trustworthiness.

\section{Discussion}
In this work, we contribute an early comparative investigation of the effectiveness and UX of multimodal interaction in an AR-based robot puppeteering system for virtual counterpart, combining LLM-assisted voice commands and hand-gesture interaction in a sequential, role-allocated architecture. An empirical evaluation has been implemented to justify our hypotheses. Our initial hypotheses—\textbf{H1}, that VG would lead to better task performance, and \textbf{H2}, that VG would enhance user experience—were not fully supported by the empirical data. Instead, both quantitative metrics and qualitative feedback consistently demonstrated that the GO condition outperformed VG across several key dimensions of performance and usability. However, we identified the advantages of multimodal interaction in several aspects.

\subsection{Reflections and Insights}
In relation to \textbf{H1}, task performance results showed that participants completed the task more efficiently in the GO condition. They successfully moved more cubes, completed the tasks in less time, and encountered fewer system interruptions. These findings reached statistical significance in key performance indicators, such as the number of cubes and TCT. The VG condition, rather than enhancing performance, introduced additional complexity. Voice commands—while conceptually beneficial for hands-free control—suffered from recognition delays and accuracy issues that disrupted task flow. The need to switch between modalities appeared to increase cognitive load and reduce operational fluency, particularly during time-sensitive or precision-based actions.

Regarding \textbf{H2}, UX and usability measures—spanning UEQ-S, NASA TLX, and custom evaluation metrics—revealed a clear preference for the GO interaction model. Participants rated GO higher in pragmatic usability, perceived effectiveness, and overall satisfaction. Although hedonic quality scores were comparable between conditions, suggesting that both interactions were similarly engaging, the practical advantages of GO were evident. The VG condition showed modest improvements in perceived temporal demand and effort (as captured in NASA TLX), yet these did not correlate with objective performance outcomes, indicating a potential disconnect between perceived and actual efficacy.

The qualitative data provided important context for interpreting these patterns. Participants frequently described the GO condition as more natural, responsive, and intuitive, reinforcing the strong usability ratings in quantitative metrics. In contrast, VG was perceived as less reliable due to occasional issues appeared on voice recognition, accent sensitivity, and a lack of fluid integration between modalities. A number of users (more on non-roboticsists' side) acknowledged voice interaction as beneficial in reducing physical effort -- especially during repetitive tasks, however, they also reported that the interruptions and mental switching detracted from overall efficiency and satisfaction.

An important caveat is that the VG condition's disadvantage may be partly attributable to the latency overhead and modality-switching cost of the voice navigation stage, rather than to the voice modality itself. Because voice was used only for the initial waypoint command while gesture handled all subsequent manipulation, our results reflect a specific role-allocation rather than a general comparison between voice and gesture.

Another concern raised on both conditions was system responsiveness. Participants emphasized the importance of clear, timely feedback and reliable control mapping. This was particularly evident in the VG condition, where inconsistencies in system response potentially undermined user trust and disrupted task continuity. These findings suggest that system predictability and feedback fidelity are critical for maintaining user confidence, especially in real-time HRI scenarios. As such, future development should prioritize seamless integration of modalities, robust input recognition, and consistent feedback loops to support more fluid and trustworthy user experiences.

\subsection{Relections from (Non-)Roboticists}
A deeper analysis of participant expertise revealed important patterns in how users interacted with and perceived the system across conditions. While GO interaction proved efficient overall, it disproportionately favored participants with prior robotics experience. Roboticists demonstrated greater control fluency, faster task execution, and fewer interruptions under GO, likely due to their familiarity with spatial manipulation and system logic. In contrast, non-roboticists were more susceptible to interaction breakdowns and exhibited lower task success under GO, indicating that the learning curve for GO may be steeper for users without technical backgrounds. However, the introduction of voice commands in the VG condition appeared to narrow this performance gap. Non-roboticists benefited from the added guidance of discrete verbal commands, which helped reduce the cognitive load associated with real-time spatial coordination. Yet for roboticists, the same addition introduced redundancy or inefficiency (For instance, the higher number of interruptions in VG), as the modality switching and system latency disrupted their already-optimized interaction strategies.

These findings suggest that multimodal AR-based HRI systems should support user-adaptive configurations that align interaction modalities with individual expertise. For example, experienced users may prefer gesture-dominant control with minimal system intervention, while novices could benefit from scaffolded interactions leveraging contextual voice commands. Designing systems that allow for such flexibility, the customizable interfaces can help accommodate a broader range of user profiles while maximizing efficiency and satisfaction for each. Additionally, further research is needed to examine how expertise influences modality switching strategies, and whether more seamless modality integration can mitigate cognitive friction for expert users.

From the UX perspective, the results indicate that our system’s core interaction design was inclusive enough to minimize large disparities between roboticists and non-roboticists. Although roboticists generally reported higher usability ratings under the GO condition, non-roboticists closed this gap under VG, particularly in some sub-metrics such as frustration level and ease of use. This suggests that the inclusion of voice input served as a compensatory mechanism, helping non-expert users feel more confident and capable in their interactions. Interestingly, workload ratings across both groups remained largely comparable, indicating that the system did not overwhelm users regardless of background, though subtle trends pointed to reduced cognitive strain for non-roboticists under VG.

These insights emphasize the value of designing interactions that scale with user expertise. While experienced users were able to adapt across modalities with relative ease, non-experts clearly benefited from the added support and structure offered by the voice interaction. This underscores a key design implication: multimodal interaction is not merely about flexibility but it plays a pivotal role in inclusivity. Even when statistical differences are absent, directional trends suggest that voice commands enhance the accessibility and engagement of HRI systems, particularly for users with limited prior experience. Future iterations should continue to explore how interaction design can dynamically support users at varying skill levels, with an emphasis on adaptive modality presentation and seamless user onboarding.

\subsection{Design Implications and Guidelines}
Building on the observed differences between user groups, this study reveals critical considerations for designing multimodal interactions in HRI systems within AR. While multimodality is often associated with increased flexibility and inclusiveness, our findings imply that such benefits are not automatically achieved through input diversity alone. The VG condition, while offered more enjoyable and intuitive experience for non-experts, introduced notable usability challenges, particularly for experienced users, due to synchronization issues between modalities, intermittent system responsiveness, and the cognitive cost of switching input modes. As mentioned previously, the effectiveness of multimodal input in the AR-HRI loop is highly contingent on user expertise, and without proper contextual integration, voice and gesture inputs can compete rather than complement each other.

Designing multimodal interaction in AR for robotics therefore \textbf{requires more than just offering multiple input options; it demands a cohesive interaction model} that adapts to users’ intent, experience level, and task complexity. Future systems should implement intelligent modality adaptation where the system can hierarchically prioritize or blend modalities based on context and user behaviors. For example, the GO condition could be foregrounded for spatial movements, while voice commands could be contextually triggered for high-level actions or when gestures interactions are not feasible. Furthermore, \textbf{multimodal systems should feature transparent feedback mechanisms} that inform users of current input states and system expectations, reducing ambiguity and improving perceived system resilience. Ultimately, \textbf{a user-centric, context-aware approach to multimodal design is essential} for achieving both performance efficiency and accessibility across varying levels of expertise in designing effective HRI systems within AR. Based on the observed use patterns, we distill five actionable guidelines with some grounded examples for designing multimodal (voice+gesture) AR robot-puppeteering systems as displayed in Table ~\ref{tab:guidelines}.

G1 operationalises expertise-aware automation level allocation \cite{parasuraman2000model,selvaggio2021autonomy}, in which experienced users benefit from lower automation (direct gesture control), while novices benefit from higher automation (parameterized voice commands) that offloads spatial planning. G2 reflects role-allocation principles in multimodal interface design \cite{oviatt1999ten,wang2024multimodal}, which assign continuous spatial manipulation to gesture and discrete tasks to speech. This is a pattern that has remained robust across numerous multimodal HRI research and is affirmed in XR-focused studies \cite{rakkolainen2021technologies,wang2025towards}. G3 and G4 correspond to the seamless transparency requirements which is important in robotic teleoperation \cite{chen2010supervisory,naughton2024integrating}, where the cost of modality switching and feedback providing must be minimized and the active control state must remain legible to the users. G5 reflects the progressive autonomy principle \cite{hebert2015supervised}, in which the balance between the user and system shifts adaptively as the user's competence develops.

\begin{table*}[t]
\centering
\footnotesize
\setlength{\tabcolsep}{4pt}
\renewcommand{\arraystretch}{1.25}

\rowcolors{3}{rowgray}{white}

\begin{tabularx}{\textwidth}{
  @{} >{\RaggedRight\arraybackslash}p{0.25cm}  
      >{\RaggedRight\arraybackslash}p{3.6cm}  
      X                                      
      X                                       
  @{}}
\toprule
\rowcolor{headergray}
\textbf{} & \textbf{Design Guideline} &
\textbf{Guideline Explanation} &
\textbf{Example} \\
\toprule
\midrule

\textbf{G1} & \textbf{Identify domain expertise and set defaults}
& Detect user expertise; start experts in \emph{Gesture-Only} (GO), novices in \emph{Voice+Gesture} (VG).
& First run: experts start in GO with gesture gain and a switch to VG; non-experts start in VG with command overlay (e.g., "move to XX", "pinch and hold") and a switch to GO. \\

\textbf{G2} & \textbf{Allocate roles to modalities}
& Use gesture for continuous, spatial control; voice for discrete, high-level, or hands-busy tasks.
& Voice: "move to red zone" for navigation. Gesture: pinch-and-drag for fine placement. \\

\textbf{G3} & \textbf{Minimize modalities switching cost}
& Enable seamless switch such as hold-to-talk; keep the context across modes; avoid re-targeting the objects.
& Hold-to-talk that does not break gesture control; no re-targeting of the object required after voice command. \\

\textbf{G4} & \textbf{Provide transparent feedback mechanism}
& Always show system state; provide clear error or progress cues.
& Colour-coded HMD border indicates active modality; audio signal confirms voice recognition; text overlay shows recognized command shortly. \\

\textbf{G5} & \textbf{Onboard progressively and adapt interaction over time}
& Begin with more guidance first; gradually reduce prompts and adjust sensitivity as users gain fluency.
& Experts start with GO; enable adjusting gesture sensitivity and showing/hiding voice settings. Non-experts start with VG with cues (e.g., visual tips); as accuracy improves, cues fade and gestures become primary.\\

\bottomrule
\end{tabularx}
\caption{Design guidelines for AR-based robot puppeteering with voice+gesture interaction.}
\label{tab:guidelines}
\end{table*}

\subsection{Generalizability and Real-World Use}
Although conducted in a controlled laboratory environment, the study’s findings have broader implications for deploying AR-mediated robotic 'puppeteer' systems in real-world contexts. The task design, which was centered on spatial manipulation, target matching, and time efficiency, reflects fundamental HRI tasks in domains such as remote inspection, field robotics, warehouse automation, or operation in hazardous environments where interacting with a virtual robot can reduce risk. The clear performance advantage of GO interaction suggests that in high-stakes or time-sensitive scenarios, streamlined unimodal systems may currently offer more reliable and efficient user control. This is particularly relevant for environments where consistency, low latency, and minimal input error are critical.

From a multimodal interaction perspective, our results highlight a tension between the theoretical benefits of multimodality and the practical costs of integrating multiple channels. Multimodal theory suggests that combining modalities can provide complementary strengths (e.g., gestures for continuous spatial control and speech for discrete, semantic commands), redundancy for error tolerance, and mutual disambiguation when signals are ambiguous. In our setup, however, the additional integration overhead, recognition errors, and temporal misalignment in VG meant that these theoretical benefits did not fully materialize and, in some cases, hindered performance. This underlines that “more modalities” are not inherently better; rather, their value depends on how well they are aligned with task demands, temporal constraints, and users’ cognitive resources.

Nevertheless, the potential of multimodal systems remains highly relevant for real-world applications, especially in situations that demand flexible user input due to physical or cognitive constraints. For instance, in hands-busy contexts such as fieldwork involving protective gear or robotic surgery, voice commands could reduce cognitive and physical strain if implemented reliably. Similarly, in visually cluttered settings, gesture recognition may offer a more dependable alternative. Generalizability, however, depends heavily on system robustness and adaptability. Factors such as fluctuating lighting conditions, limited user training, and unexpected physical interruptions must be considered in future deployment. This underscores the importance of building multimodal systems that are resilient to environmental variability and personalized to user needs. Our findings suggest that while gesture-only systems may serve as a strong foundation, the long-term success of multimodal interaction in real-world HRI will rely on resolving current integration challenges and fostering adaptive, user-aware design paradigms that operationalize multimodal principles in a task- and context-sensitive manner.

\subsection{Limitations and Future Work}
While the study yielded valuable insights into the comparative effectiveness of GO and VG interactions, several limitations should be acknowledged. One key limitation concerns the technical maturity of the system, particularly the voice command implementation. The VG condition relied on an LLM-assisted speech processing module which, although offering semantic richness, still exhibited latency, inconsistency in recognition, and limited robustness to accents. Participants occasionally experienced breakdowns in interaction flow due to delayed or unrecognized voice inputs. These issues likely affected both subjective workload and task performance. In its current early-stage form, the voice interaction pipeline lacks the speed and reliability required for precise, time-critical robotic control tasks, especially those demanding immediate feedback.

Although the gesture interface was generally perceived as intuitive, it also suffered from sensitivity issues and occasional misinterpretations, particularly during prolonged or rapid interactions. The gesture recognition system did not always respond fluidly to user input, and the robot’s inherent fragility (e.g., when users moved their hands quickly) led to safety interruptions that disrupted task execution and undermined user trust. Furthermore, the current task design is relatively monotonous, and voice interaction was only used in the first task stage, which may have influenced both task performance and user perceptions of the VG condition.

A related methodological limitation concerns the 300s time limit on the task execution of each trial. This was a deliberate choice to control the total study length and to mitigate fatigue effects from prolonged AR HMD use. However, a substantial proportion of VG trials reached the ceiling while most GO trials did not, partially due to the time spent repeating unrecognized voice commands (as reported by some participants) was included in the recorded duration, which may affect the performance results. Hence, the reported results of TCT should be read as conservative lower bounds on the true GO–VG difference; the same caveat applies to the roboticist vs. non-roboticist comparison. Future studies could address this by using a longer time duration with mandatory breaks, reporting more robust performance measures (e.g., cubes per minute).

A further limitation is the absence of baseline conditions using established teleoperation interfaces such as joysticks, touchscreens, or 6-DoF controllers. As a result, our findings should be interpreted strictly as relative differences between GO and VG within our AR 'puppeteer' context, rather than as evidence that either modality outperforms conventional teleoperation tools. Future work should embed our AR 'puppeteer' interface into a broader comparative framework, systematically evaluating its strengths and weaknesses against these widely used baselines.

In addition, the design implications and guidelines we derive are constrained by the scope of this study. They are based on a single AR 'puppeteer' system, a single pick-and-place teleoperation task, and two specific modality configurations (GO and VG) with a particular participant sample. Consequently, they should be used as actionable recommendations for similar AR-based robotic teleoperation scenarios, rather than definitive or universal rules for multimodal HRI.

From a design perspective, future work should address the limitations of both input modalities. The voice command techniques could benefit from a more robust and adaptive recognition engine, potentially fine-tuned to domain-specific vocabularies and augmented with real-time clarification mechanisms for ambiguous queries. For gesture input, improvements in gesture-to-action mappings, more resilient tracking algorithms, and clearer visual or haptic feedback could mitigate usability issues and reduce interruptions during rapid movement. Beyond improving the underlying technology, future studies should explore richer task designs that engage voice and gesture across multiple stages and roles, examining different modality combinations and division of labor over time. Finally, expanding evaluation beyond a controlled lab setting and diversifying task scenarios will be essential for assessing the generalizability of AR-based puppeteering to real-world HRI applications.

\section{Conclusion}
This study explored the impact of gesture-only versus multimodal (VG) interactions in an AR-based robot puppeteering system. Results showed that GO consistently outperformed VG in task performance, usability, and user satisfaction. Voice input benefited novice users, narrowing the performance gap with experts; however, it introduced cognitive overhead and system inefficiencies that affected overall UX. As an early-stage exploratory study, our findings highlight the importance of optimizing modality integration (e.g., the sequential role-allocated multimodality adopted in this paper) and suggest that hand gesture-based interaction remains a robust solution for robot 'puppeteer' metaphoric system in AR -- especially for domain experts -- while multimodal designs should be tailored to users’ expertise. Future work should explore adaptive, user-aware multimodal systems that balance flexibility with reliability across diverse user contexts.

\backmatter

\bmhead{Acknowledgements}

Acknowledgements are not compulsory. Where included they should be brief. Grant or contribution numbers may be acknowledged.

Please refer to Journal-level guidance for any specific requirements.

\section*{Declarations}

\begin{itemize}
\item Funding
\item Conflict of interest/Competing interests (check journal-specific guidelines for which heading to use)
\item Ethics approval and consent to participate
\item Consent for publication
\item Data availability 
\item Materials availability
\item Code availability 
\item Author contribution
\end{itemize}








\bibliography{sn-bibliography}

@article{ribeiro2021robotic,
  title={Robotic process automation and artificial intelligence in industry 4.0--a literature review},
  author={Ribeiro, Jorge and Lima, Rui and Eckhardt, Tiago and Paiva, Sara},
  journal={Procedia Computer Science},
  volume={181},
  pages={51--58},
  year={2021},
  publisher={Elsevier}
}

@article{hentout2019human,
  title={Human--robot interaction in industrial collaborative robotics: a literature review of the decade 2008--2017},
  author={Hentout, Abdelfetah and Aouache, Mustapha and Maoudj, Abderraouf and Akli, Isma},
  journal={Advanced Robotics},
  volume={33},
  number={15-16},
  pages={764--799},
  year={2019},
  publisher={Taylor \& Francis}
}

@inproceedings{richert2016socializing,
  title={Socializing with robots: Human-robot interactions within a virtual environment},
  author={Richert, Anja and Shehadeh, Mohammad A and M{\"u}ller, Sarah L and Schr{\"o}der, Stefan and Jeschke, Sabina},
  booktitle={2016 IEEE workshop on advanced robotics and its social impacts (ARSO)},
  pages={49--54},
  year={2016},
  organization={IEEE}
}

@article{olaronke2017state,
  title={State of the art: a study of human-robot interaction in healthcare},
  author={Olaronke, Iroju and Oluwaseun, Ojerinde and Rhoda, Ikono},
  journal={International Journal of Information Engineering and Electronic Business},
  volume={9},
  number={3},
  pages={43},
  year={2017},
  publisher={Modern Education and Computer Science Press}
}

@article{hopko2021effect,
  title={Effect of cognitive fatigue, operator sex, and robot assistance on task performance metrics, workload, and situation awareness in human-robot collaboration},
  author={Hopko, Sarah K and Khurana, Riya and Mehta, Ranjana K and Pagilla, Prabhakar R},
  journal={IEEE Robotics and Automation Letters},
  volume={6},
  number={2},
  pages={3049--3056},
  year={2021},
  publisher={IEEE}
}

@article{apraiz2023evaluation,
  title={Evaluation of user experience in human--robot interaction: a systematic literature review},
  author={Apraiz, Ainhoa and Lasa, Ganix and Mazmela, Maitane},
  journal={International Journal of Social Robotics},
  volume={15},
  number={2},
  pages={187--210},
  year={2023},
  publisher={Springer}
}

@inproceedings{zhou2008trends,
  title={Trends in augmented reality tracking, interaction and display: A review of ten years of ISMAR},
  author={Zhou, Feng and Duh, Henry Been-Lirn and Billinghurst, Mark},
  booktitle={2008 7th IEEE/ACM International Symposium on Mixed and Augmented Reality},
  pages={193--202},
  year={2008},
  organization={IEEE}
}

@article{saeedi2018navigating,
  title={Navigating the landscape for real-time localization and mapping for robotics and virtual and augmented reality},
  author={Saeedi, Sajad and Bodin, Bruno and Wagstaff, Harry and Nisbet, Andy and Nardi, Luigi and Mawer, John and Melot, Nicolas and Palomar, Oscar and Vespa, Emanuele and Spink, Tom and others},
  journal={Proceedings of the IEEE},
  volume={106},
  number={11},
  pages={2020--2039},
  year={2018},
  publisher={IEEE}
}

@inproceedings{suzuki2022augmented,
  title={Augmented reality and robotics: A survey and taxonomy for ar-enhanced human-robot interaction and robotic interfaces},
  author={Suzuki, Ryo and Karim, Adnan and Xia, Tian and Hedayati, Hooman and Marquardt, Nicolai},
  booktitle={Proceedings of the 2022 CHI Conference on Human Factors in Computing Systems},
  pages={1--33},
  year={2022}
}

@article{yoo2025study,
  title={A study on view sharing AR interface for improving situation awareness during military operations},
  author={Yoo, Gi Sung and Ji, Yong Gu},
  journal={International Journal of Human--Computer Interaction},
  volume={41},
  number={4},
  pages={2211--2226},
  year={2025},
  publisher={Taylor \& Francis}
}

@inproceedings{bambusek2019combining,
  title={Combining interactive spatial augmented reality with head-mounted display for end-user collaborative robot programming},
  author={Bambu{\^s}ek, Daniel and Materna, Zden{\v{e}}k and Kapinus, Michal and Beran, V{\'\i}t{\v{e}}zslav and Smr{\v{z}}, Pavel},
  booktitle={2019 28th IEEE international conference on robot and human interactive communication (RO-MAN)},
  pages={1--8},
  year={2019},
  organization={IEEE}
}

@inproceedings{gaschler2014intuitive,
  title={Intuitive robot tasks with augmented reality and virtual obstacles},
  author={Gaschler, Andre and Springer, Maximilian and Rickert, Markus and Knoll, Alois},
  booktitle={2014 IEEE International Conference on Robotics and Automation (ICRA)},
  pages={6026--6031},
  year={2014},
  organization={IEEE}
}

@article{chong2009robot,
  title={Robot programming using augmented reality: An interactive method for planning collision-free paths},
  author={Chong, Jonathan Wun Shiung and Ong, SKc and Nee, Andrew YC and Youcef-Youmi, KB},
  journal={Robotics and Computer-Integrated Manufacturing},
  volume={25},
  number={3},
  pages={689--701},
  year={2009},
  publisher={Elsevier}
}

@inproceedings{lunding2023ar,
  title={Ar-supported human-robot collaboration: Facilitating workspace awareness and parallelized assembly tasks},
  author={Lunding, Rasmus S and Lystb{\ae}k, Mathias N and Feuchtner, Tiare and Gr{\o}nb{\ae}k, Kaj},
  booktitle={2023 IEEE International Symposium on Mixed and Augmented Reality (ISMAR)},
  pages={1064--1073},
  year={2023},
  organization={IEEE}
}

@inproceedings{walker2018communicating,
  title={Communicating robot motion intent with augmented reality},
  author={Walker, Michael and Hedayati, Hooman and Lee, Jennifer and Szafir, Daniel},
  booktitle={Proceedings of the 2018 ACM/IEEE International Conference on Human-Robot Interaction},
  pages={316--324},
  year={2018}
}

@article{moya2023augmented,
  title={Augmented reality for supporting workers in human--robot collaboration},
  author={Moya, Ana and Bastida, Leire and Aguirrezabal, Pablo and Pantano, Matteo and Abril-Jim{\'e}nez, Patricia},
  journal={Multimodal Technologies and Interaction},
  volume={7},
  number={4},
  pages={40},
  year={2023},
  publisher={MDPI}
}

@inproceedings{arevalo2021assisting,
  title={Assisting manipulation and grasping in robot teleoperation with augmented reality visual cues},
  author={Arevalo Arboleda, Stephanie and R{\"u}cker, Franziska and Dierks, Tim and Gerken, Jens},
  booktitle={Proceedings of the 2021 CHI conference on human factors in computing systems},
  pages={1--14},
  year={2021}
}

@article{makhataeva2020augmented,
  title={Augmented reality for robotics: A review},
  author={Makhataeva, Zhanat and Varol, Huseyin Atakan},
  journal={Robotics},
  volume={9},
  number={2},
  pages={21},
  year={2020},
  publisher={MDPI}
}

@inproceedings{mohareri2011autonomous,
  title={Autonomous humanoid robot navigation using augmented reality technique},
  author={Mohareri, Omid and Rad, Ahmad B},
  booktitle={2011 IEEE International Conference on Mechatronics},
  pages={463--468},
  year={2011},
  organization={IEEE}
}

@inproceedings{barakonyi2004agents,
  title={Agents that talk and hit back: Animated agents in augmented reality},
  author={Barakonyi, Istv{\'a}n and Psik, Thomas and Schmalstieg, Dieter},
  booktitle={Third IEEE and ACM International Symposium on Mixed and Augmented Reality},
  pages={141--150},
  year={2004},
  organization={IEEE}
}

@inproceedings{van2024puppeteer,
  title={Puppeteer your robot: Augmented reality leader-follower teleoperation},
  author={Van Haastregt, Jonne and Welle, Michael C and Zhang, Yuchong and Kragic, Danica},
  booktitle={2024 IEEE-RAS 23rd International Conference on Humanoid Robots (Humanoids)},
  pages={1019--1026},
  year={2024},
  organization={IEEE}
}

@inproceedings{salem2011friendly,
  title={A friendly gesture: Investigating the effect of multimodal robot behavior in human-robot interaction},
  author={Salem, Maha and Rohlfing, Katharina and Kopp, Stefan and Joublin, Frank},
  booktitle={2011 ro-man},
  pages={247--252},
  year={2011},
  organization={IEEE}
}

@article{torta2015evaluation,
  title={Evaluation of unimodal and multimodal communication cues for attracting attention in human--robot interaction},
  author={Torta, Elena and van Heumen, Jim and Piunti, Francesco and Romeo, Luca and Cuijpers, Raymond},
  journal={International Journal of Social Robotics},
  volume={7},
  pages={89--96},
  year={2015},
  publisher={Springer}
}

@article{okita2011multimodal,
  title={Multimodal approach to affective human-robot interaction design with children},
  author={Okita, Sandra Y and Ng-Thow-Hing, Victor and Sarvadevabhatla, Ravi K},
  journal={ACM Transactions on Interactive Intelligent Systems (TiiS)},
  volume={1},
  number={1},
  pages={1--29},
  year={2011},
  publisher={ACM New York, NY, USA}
}

@inproceedings{ali2019design,
  title={Design of seamless multi-modal interaction framework for intelligent virtual agents in wearable mixed reality environment},
  author={Ali, Ghazanfar and Le, Hong-Quan and Kim, Junho and Hwang, Seung-Won and Hwang, Jae-In},
  booktitle={Proceedings of the 32nd International Conference on Computer Animation and Social Agents},
  pages={47--52},
  year={2019}
}

@article{oviatt1999ten,
  title={Ten myths of multimodal interaction},
  author={Oviatt, Sharon},
  journal={Communications of the ACM},
  volume={42},
  number={11},
  pages={74--81},
  year={1999},
  publisher={ACM New York, NY, USA}
}

@article{lee2013usability,
  title={A usability study of multimodal input in an augmented reality environment},
  author={Lee, Minkyung and Billinghurst, Mark and Baek, Woonhyuk and Green, Richard and Woo, Woontack},
  journal={Virtual Reality},
  volume={17},
  pages={293--305},
  year={2013},
  publisher={Springer}
}

@article{lunghi2019multimodal,
  title={Multimodal human-robot interface for accessible remote robotic interventions in hazardous environments},
  author={Lunghi, Giacomo and Marin, Raul and Di Castro, Mario and Masi, Alessandro and Sanz, Pedro J},
  journal={IEEE Access},
  volume={7},
  pages={127290--127319},
  year={2019},
  publisher={IEEE}
}

@article{lazzeri2014development,
  title={Development and testing of a multimodal acquisition platform for human-robot interaction affective studies},
  author={Lazzeri, Nicole and Mazzei, Daniele and De Rossi, Danilo},
  journal={Journal of Human-Robot Interaction},
  volume={3},
  number={2},
  pages={1--24},
  year={2014},
  publisher={Journal of Human-Robot Interaction Steering Committee}
}

@article{cao2023investigating,
  title={Investigating the role of multi-modal social cues in human-robot collaboration in industrial settings},
  author={Cao, Hoang-Long and Scholz, Constantin and De Winter, Joris and Makrini, Ilias El and Vanderborght, Bram},
  journal={International Journal of Social Robotics},
  volume={15},
  number={7},
  pages={1169--1179},
  year={2023},
  publisher={Springer}
}

@inproceedings{williams2018virtual,
  title={Virtual, augmented, and mixed reality for human-robot interaction},
  author={Williams, Tom and Szafir, Daniel and Chakraborti, Tathagata and Ben Amor, Heni},
  booktitle={Companion of the 2018 ACM/IEEE International Conference on Human-Robot Interaction},
  pages={403--404},
  year={2018}
}

@article{freund1999projective,
  title={Projective virtual reality: Bridging the gap between virtual reality and robotics},
  author={Freund, Eckhard and Rossmann, Juergen},
  journal={IEEE transactions on robotics and automation},
  volume={15},
  number={3},
  pages={411--422},
  year={1999},
  publisher={IEEE}
}

@inproceedings{villani2018use,
  title={Use of virtual reality for the evaluation of human-robot interaction systems in complex scenarios},
  author={Villani, Valeria and Capelli, Beatrice and Sabattini, Lorenzo},
  booktitle={2018 27th IEEE international symposium on robot and human interactive communication (RO-MAN)},
  pages={422--427},
  year={2018},
  organization={IEEE}
}

@article{matsas2017design,
  title={Design of a virtual reality training system for human--robot collaboration in manufacturing tasks},
  author={Matsas, Elias and Vosniakos, George-Christopher},
  journal={International Journal on Interactive Design and Manufacturing (IJIDeM)},
  volume={11},
  pages={139--153},
  year={2017},
  publisher={Springer}
}

@inproceedings{murnane2021simulator,
  title={A simulator for human-robot interaction in virtual reality},
  author={Murnane, Mark and Higgins, Padraig and Saraf, Monali and Ferraro, Francis and Matuszek, Cynthia and Engel, Don},
  booktitle={2021 IEEE Conference on Virtual Reality and 3D User Interfaces Abstracts and Workshops (VRW)},
  pages={470--471},
  year={2021},
  organization={IEEE}
}

@inproceedings{walker2019robot,
  title={Robot teleoperation with augmented reality virtual surrogates},
  author={Walker, Michael E and Hedayati, Hooman and Szafir, Daniel},
  booktitle={2019 14th ACM/IEEE International Conference on Human-Robot Interaction (HRI)},
  pages={202--210},
  year={2019},
  organization={IEEE}
}

@inproceedings{hedayati2018improving,
  title={Improving collocated robot teleoperation with augmented reality},
  author={Hedayati, Hooman and Walker, Michael and Szafir, Daniel},
  booktitle={Proceedings of the 2018 ACM/IEEE International Conference on Human-Robot Interaction},
  pages={78--86},
  year={2018}
}

@inproceedings{chacko2019augmented,
  title={An augmented reality interface for human-robot interaction in unconstrained environments},
  author={Chacko, Sonia Mary and Kapila, Vikram},
  booktitle={2019 IEEE/RSJ International Conference on Intelligent Robots and Systems (IROS)},
  pages={3222--3228},
  year={2019},
  organization={IEEE}
}

@article{michalos2016augmented,
  title={Augmented reality (AR) applications for supporting human-robot interactive cooperation},
  author={Michalos, George and Karagiannis, Panagiotis and Makris, Sotiris and Tok{\c{c}}alar, {\"O}nder and Chryssolouris, George},
  journal={Procedia CIRP},
  volume={41},
  pages={370--375},
  year={2016},
  publisher={Elsevier}
}

@article{szczurek2023multimodal,
  title={Multimodal multi-user mixed reality human--robot interface for remote operations in hazardous environments},
  author={Szczurek, Krzysztof Adam and Prades, Raul Marin and Matheson, Eloise and Rodriguez-Nogueira, Jose and Di Castro, Mario},
  journal={IEEE Access},
  volume={11},
  pages={17305--17333},
  year={2023},
  publisher={IEEE}
}

@incollection{chan2022multimodal,
  title={A multimodal system using augmented reality, gestures, and tactile feedback for robot trajectory programming and execution},
  author={Chan, Wesley P and Quintero, Camilo Perez and Pan, Matthew KXJ and Sakr, Maram and Van der Loos, HF Machiel and Croft, Elizabeth},
  booktitle={Virtual Reality},
  pages={142--158},
  year={2022},
  publisher={River Publishers}
}

@article{park2021hands,
  title={Hands-free human--robot interaction using multimodal gestures and deep learning in wearable mixed reality},
  author={Park, Kyeong-Beom and Choi, Sung Ho and Lee, Jae Yeol and Ghasemi, Yalda and Mohammed, Mustafa and Jeong, Heejin},
  journal={IEEE Access},
  volume={9},
  pages={55448--55464},
  year={2021},
  publisher={IEEE}
}

@article{savage1998virbot,
  title={The VirBot: a virtual reality robot driven with multimodal commands},
  author={Savage-Carmona, Jesus and Billinghurst, Mark and Holden, Alistair},
  journal={Expert Systems with Applications},
  volume={15},
  number={3-4},
  pages={413--419},
  year={1998},
  publisher={Elsevier}
}

@article{polvi2017handheld,
  title={Handheld guides in inspection tasks: Augmented reality versus picture},
  author={Polvi, Jarkko and Taketomi, Takafumi and Moteki, Atsunori and Yoshitake, Toshiyuki and Fukuoka, Toshiyuki and Yamamoto, Goshiro and Sandor, Christian and Kato, Hirokazu},
  journal={IEEE transactions on visualization and computer graphics},
  volume={24},
  number={7},
  pages={2118--2128},
  year={2017},
  publisher={IEEE}
}

@inproceedings{shayesteh2021investigating,
  title={Investigating the impact of construction robots autonomy level on workers' cognitive load},
  author={Shayesteh, Shayan and Jebelli, Houtan},
  booktitle={Canadian Society of Civil Engineering Annual Conference},
  pages={255--267},
  year={2021},
  organization={Springer}
}

@article{pick2016design,
  title={Design and evaluation of data annotation workflows for cave-like virtual environments},
  author={Pick, Sebastian and Weyers, Benjamin and Hentschel, Bernd and Kuhlen, Torsten W},
  journal={IEEE transactions on visualization and computer graphics},
  volume={22},
  number={4},
  pages={1452--1461},
  year={2016},
  publisher={IEEE}
}

@inproceedings{lin2024usability,
  title={Usability, Acceptance, and Trust of Privacy Protection Mechanisms and Identity Management in Social Virtual Reality},
  author={Lin, Jinghuai and Rack, Christian and Wienrich, Carolin and Latoschik, Marc Erich},
  booktitle={2024 IEEE International Symposium on Mixed and Augmented Reality (ISMAR)},
  pages={130--139},
  year={2024},
  organization={IEEE}
}

@inproceedings{chang2024perceived,
  title={Perceived Empathy in Mixed Reality: Assessing the Impact of Empathic Agents' Awareness of User Physiological States},
  author={Chang, Zhuang and Kim, Kangsoo and Gupta, Kunal and Abouelenin, Jamila and Xiao, Zirui and Gu, Boyang and Bai, Huidong and Billinghurst, Mark},
  booktitle={2024 IEEE International Symposium on Mixed and Augmented Reality (ISMAR)},
  pages={406--415},
  year={2024},
  organization={IEEE}
}

@inproceedings{zhao2024glanxr,
  title={GlanXR: A Hands-Free Fast Switching System for Virtual Screens},
  author={Zhao, Guanghan and Orlosky, Jason and Kiyokawa, Kiyoshi and Uranishi, Yuki},
  booktitle={2024 IEEE International Symposium on Mixed and Augmented Reality (ISMAR)},
  pages={111--119},
  year={2024},
  organization={IEEE}
}

@article{schrepp2017design,
  title={Design and Evaluation of a Short Version of the User Experience Questionnaire (UEQ-S).},
  author={Schrepp, Martin and Hinderks, Andreas and Thomaschewski, J{\"o}rg},
  journal={International Journal of Interactive Multimedia \& Artificial Intelligence},
  volume={4},
  number={6},
  year={2017}
}

@article{fang2014novel,
  title={Novel AR-based interface for human-robot interaction and visualization},
  author={Fang, HC and Ong, Soh-Khim and Nee, Andrew YC},
  journal={Advances in Manufacturing},
  volume={2},
  pages={275--288},
  year={2014},
  publisher={Springer}
}

@inproceedings{ostanin2020human,
  title={Human-robot interaction for robotic manipulator programming in Mixed Reality},
  author={Ostanin, Mikhail and Mikhel, Stanislav and Evlampiev, Alexey and Skvortsova, Valeria and Klimchik, Alexandr},
  booktitle={2020 IEEE international conference on robotics and automation (ICRA)},
  pages={2805--2811},
  year={2020},
  organization={IEEE}
}

@article{pandey2014towards,
  title={Towards human-level semantics understanding of human-centered object manipulation tasks for HRI: reasoning about effect, ability, effort and perspective taking},
  author={Pandey, Amit Kumar and Alami, Rachid},
  journal={International Journal of Social Robotics},
  volume={6},
  pages={593--620},
  year={2014},
  publisher={Springer}
}

@inproceedings{seaborn2010exploring,
  title={Exploring the interplay of visual and haptic modalities in a pattern-matching task},
  author={Seaborn, Katie and Riecke, Bernhard E and Antle, Alissa N},
  booktitle={2010 IEEE International Symposium on Haptic Audio Visual Environments and Games},
  pages={1--6},
  year={2010},
  organization={IEEE}
}

@article{swan2015matching,
  title={Matching and reaching depth judgments with real and augmented reality targets},
  author={Swan, J Edward and Singh, Gurjot and Ellis, Stephen R},
  journal={IEEE transactions on visualization and computer graphics},
  volume={21},
  number={11},
  pages={1289--1298},
  year={2015},
  publisher={IEEE}
}

@article{nickerson1973visual,
  title={Visual pattern matching: an investigation of some effects of decision task, auditory codability, and spatial correspondence.},
  author={Nickerson, RS and Pew, RW},
  journal={Journal of Experimental Psychology},
  volume={98},
  number={1},
  pages={36},
  year={1973},
  publisher={American Psychological Association}
}

@inproceedings{wang2024eve,
  title={EVE: Enabling anyone to train robots using augmented reality},
  author={Wang, Jun and Chang, Chun-Cheng and Duan, Jiafei and Fox, Dieter and Krishna, Ranjay},
  booktitle={Proceedings of the 37th Annual ACM Symposium on User Interface Software and Technology},
  pages={1--13},
  year={2024}
}

@inproceedings{duan2023ar2,
  title={AR2-D2: Training a Robot Without a Robot},
  author={Duan, Jiafei and Wang, Yi Ru and Shridhar, Mohit and Fox, Dieter and Krishna, Ranjay},
  booktitle={Conference on Robot Learning},
  pages={2838--2848},
  year={2023},
  organization={PMLR}
}

@inproceedings{zhang2025llm,
  title={LLM-Driven Augmented Reality Puppeteer: Controller-Free Voice-Commanded Robot Teleoperation},
  author={Zhang, Yuchong and Orthmann, Bastian and Welle, Michael C and Van Haastregt, Jonne and Kragic, Danica},
  booktitle={International Conference on Human-Computer Interaction},
  pages={97--112},
  year={2025},
  organization={Springer}
}

@inproceedings{krupke2018comparison,
  title={Comparison of multimodal heading and pointing gestures for co-located mixed reality human-robot interaction},
  author={Krupke, Dennis and Steinicke, Frank and Lubos, Paul and Jonetzko, Yannick and G{\"o}rner, Michael and Zhang, Jianwei},
  booktitle={2018 IEEE/RSJ International Conference on Intelligent Robots and Systems (IROS)},
  pages={1--9},
  year={2018},
  organization={IEEE}
}

@inproceedings{zhang2023see,
  title={See or hear? exploring the effect of visual/audio hints and gaze-assisted instant post-task feedback for visual search tasks in ar},
  author={Zhang, Yuchong and Nowak, Adam and Xuan, Yueming and Romanowski, Andrzej and Fjeld, Morten},
  booktitle={2023 IEEE International Symposium on Mixed and Augmented Reality (ISMAR)},
  pages={1113--1122},
  year={2023},
  organization={IEEE}
}

@article{zhang2021supporting,
  title={Supporting visualization analysis in industrial process tomography by using augmented reality—a case study of an industrial microwave drying system},
  author={Zhang, Yuchong and Omrani, Adel and Yadav, Rahul and Fjeld, Morten},
  journal={Sensors},
  volume={21},
  number={19},
  pages={6515},
  year={2021},
  publisher={MDPI}
}

@inproceedings{nowak2021augmented,
  title={Augmented reality with industrial process tomography: to support complex data analysis in 3D space},
  author={Nowak, Adam and Zhang, Yuchong and Romanowski, Andrzej and Fjeld, Morten},
  booktitle={Adjunct Proceedings of the 2021 ACM International Joint Conference on Pervasive and Ubiquitous Computing and Proceedings of the 2021 ACM International Symposium on Wearable Computers},
  pages={56--58},
  year={2021}
}

@article{xia2024shaping,
  title={Shaping high-performance wearable robots for human motor and sensory reconstruction and enhancement},
  author={Xia, Haisheng and Zhang, Yuchong and Rajabi, Nona and Taleb, Farzaneh and Yang, Qunting and Kragic, Danica and Li, Zhijun},
  journal={Nature Communications},
  volume={15},
  number={1},
  pages={1760},
  year={2024},
  publisher={Nature Publishing Group UK London}
}

@inproceedings{zhang2024vision,
  title={Vision beyond boundaries: An initial design space of domain-specific large vision models in human-robot interaction},
  author={Zhang, Yuchong and Ma, Yong and Kragic, Danica},
  booktitle={Adjunct Proceedings of the 26th International Conference on Mobile Human-Computer Interaction},
  pages={1--8},
  year={2024}
}

@article{zhang2025mind,
  title={Mind meets robots: a review of EEG-based brain-robot interaction systems},
  author={Zhang, Yuchong and Rajabi, Nona and Taleb, Farzaneh and Matviienko, Andrii and Ma, Yong and Bj{\"o}rkman, M{\aa}rten and Kragic, Danica},
  journal={International Journal of Human--Computer Interaction},
  pages={1--32},
  year={2025},
  publisher={Taylor \& Francis}
}

@inproceedings{zhang2023playing,
  title={Playing with data: An augmented reality approach to interact with visualizations of industrial process tomography},
  author={Zhang, Yuchong and Xuan, Yueming and Yadav, Rahul and Omrani, Adel and Fjeld, Morten},
  booktitle={IFIP Conference on Human-Computer Interaction},
  pages={123--144},
  year={2023},
  organization={Springer}
}

@inproceedings{zhang2022initial,
  title={An initial exploration of visual cues in head-mounted display augmented reality for book searching},
  author={Zhang, Yuchong and Nowak, Adam and Romanowski, Andrzej and Fjeld, Morten},
  booktitle={Proceedings of the 21st International Conference on Mobile and Ubiquitous Multimedia},
  pages={273--275},
  year={2022}
}

@inproceedings{zhang2021novel,
  title={A novel augmented reality system to support volumetric visualization in industrial process tomography},
  author={Zhang, Yuchong and Yadav, Rahul and Omrani, Adel and Fjeld, Morten},
  booktitle={IHCI GET 2021 Proceedings. Proceedings of the IADIS International Conference Interfaces and Human Computer Interaction Proceedings of the IADIS International Conference Game and Entertainment Technologies. Hrsg.: K. Blashki},
  pages={3},
  year={2021}
}

@inproceedings{zhang2024human,
  title={Human-centered AI technologies in human-robot interaction for social settings},
  author={Zhang, Yuchong and Kassem, Khaled and Gong, Zhengya and Mo, Fan and Ma, Yong and Kirjavainen, Emma and H{\"a}kkil{\"a}, Jonna},
  booktitle={Proceedings of the International Conference on Mobile and Ubiquitous Multimedia},
  pages={501--505},
  year={2024}
}

@article{kyrarini2021survey,
  title={A survey of robots in healthcare},
  author={Kyrarini, Maria and Lygerakis, Fotios and Rajavenkatanarayanan, Akilesh and Sevastopoulos, Christos and Nambiappan, Harish Ram and Chaitanya, Kodur Krishna and Babu, Ashwin Ramesh and Mathew, Joanne and Makedon, Fillia},
  journal={Technologies},
  volume={9},
  number={1},
  pages={8},
  year={2021},
  publisher={MDPI}
}

@article{turk2014multimodal,
  title={Multimodal interaction: A review},
  author={Turk, Matthew},
  journal={Pattern recognition letters},
  volume={36},
  pages={189--195},
  year={2014},
  publisher={Elsevier}
}

@article{parasuraman2000model,
  title={A model for types and levels of human interaction with automation},
  author={Parasuraman, Raja and Sheridan, Thomas B and Wickens, Christopher D},
  journal={IEEE Transactions on systems, man, and cybernetics-Part A: Systems and Humans},
  volume={30},
  number={3},
  pages={286--297},
  year={2000},
  publisher={IEEE}
}

@article{biener2022quantifying,
  title={Quantifying the effects of working in VR for one week},
  author={Biener, Verena and Kalamkar, Snehanjali and Nouri, Negar and Ofek, Eyal and Pahud, Michel and Dudley, John J and Hu, Jinghui and Kristensson, Per Ola and Weerasinghe, Maheshya and Pucihar, Klen {\v{C}}opi{\v{c}} and others},
  journal={IEEE Transactions on Visualization and Computer Graphics},
  volume={28},
  number={11},
  pages={3810--3820},
  year={2022},
  publisher={IEEE}
}

@article{chen2010supervisory,
  title={Supervisory control of multiple robots: Human-performance issues and user-interface design},
  author={Chen, Jessie YC and Barnes, Michael J and Harper-Sciarini, Michelle},
  journal={IEEE Transactions on Systems, Man, and Cybernetics, Part C (Applications and Reviews)},
  volume={41},
  number={4},
  pages={435--454},
  year={2010},
  publisher={IEEE}
}

@inproceedings{naughton2024integrating,
  title={Integrating open-world shared control in immersive avatars},
  author={Naughton, Patrick and Nam, James Seungbum and Stratton, Andrew and Hauser, Kris},
  booktitle={2024 IEEE International Conference on Robotics and Automation (ICRA)},
  pages={17807--17813},
  year={2024},
  organization={IEEE}
}

@article{selvaggio2021autonomy,
  title={Autonomy in physical human-robot interaction: A brief survey},
  author={Selvaggio, Mario and Cognetti, Marco and Nikolaidis, Stefanos and Ivaldi, Serena and Siciliano, Bruno},
  journal={IEEE Robotics and Automation Letters},
  volume={6},
  number={4},
  pages={7989--7996},
  year={2021},
  publisher={IEEE}
}

@article{wang2024multimodal,
  title={Multimodal human--robot interaction for human-centric smart manufacturing: a survey},
  author={Wang, Tian and Zheng, Pai and Li, Shufei and Wang, Lihui},
  journal={Advanced Intelligent Systems},
  volume={6},
  number={3},
  pages={2300359},
  year={2024},
  publisher={Wiley Online Library}
}

@article{wang2025towards,
  title={Towards spatial computing: recent advances in multimodal natural interaction for Extended Reality headsets},
  author={Wang, Zhi-Min and Rao, Mao-Hang and Ye, Shang-Hua and Song, Wei-Tao and Lu, Feng},
  journal={Frontiers of Computer Science},
  volume={19},
  number={12},
  pages={1912708},
  year={2025},
  publisher={Springer}
}

@article{rakkolainen2021technologies,
  title={Technologies for multimodal interaction in extended reality—a scoping review},
  author={Rakkolainen, Ismo and Farooq, Ahmed and Kangas, Jari and Hakulinen, Jaakko and Rantala, Jussi and Turunen, Markku and Raisamo, Roope},
  journal={Multimodal Technologies and Interaction},
  volume={5},
  number={12},
  pages={81},
  year={2021},
  publisher={MDPI}
}

@inproceedings{hebert2015supervised,
  title={Supervised remote robot with guided autonomy and teleoperation (SURROGATE): a framework for whole-body manipulation},
  author={Hebert, Paul and Ma, Jeremy and Borders, James and Aydemir, Alper and Bajracharya, Max and Hudson, Nicolas and Shankar, Krishna and Karumanchi, Sisir and Douillard, Bertrand and Burdick, Joel},
  booktitle={2015 IEEE international conference on robotics and automation (ICRA)},
  pages={5509--5516},
  year={2015},
  organization={IEEE}
}

@inproceedings{tao2025lams,
  title={Lams: Llm-driven automatic mode switching for assistive teleoperation},
  author={Tao, Yiran and Yang, Jehan and Ding, Dan and Erickson, Zackory},
  booktitle={2025 20th ACM/IEEE International Conference on Human-Robot Interaction (HRI)},
  pages={242--251},
  year={2025},
  organization={IEEE}
}

@inproceedings{hu2025gesprompt,
  title={GesPrompt: Leveraging Co-Speech Gestures to Augment LLM-Based Interaction in Virtual Reality},
  author={Hu, Xiyun and Ma, Dizhi and He, Fengming and Zhu, Zhengzhe and Hsia, Shao-Kang and Zhu, Chenfei and Liu, Ziyi and Ramani, Karthik},
  booktitle={Proceedings of the 2025 ACM Designing Interactive Systems Conference},
  pages={59--80},
  year={2025}
}

@article{wang2024towards,
  title={Towards massive interaction with generalist robotics: a systematic review of XR-enabled remote human-robot interaction systems},
  author={Wang, Xian and Shen, Luyao and Lee, Lik-Hang},
  journal={arXiv preprint arXiv:2403.11384},
  year={2024}
}










\end{document}